\documentclass{article}
\pdfoutput=1 
\usepackage{jheppub}
\usepackage{float}
\usepackage{ulem}
\usepackage{mathrsfs}
\usepackage{natbib}
\usepackage{tikz}
\usepackage{setspace}
\usepackage{amsfonts}
\usepackage{amssymb}
\usepackage{amsmath}
\usepackage{esint}       
\usepackage{epsfig}
\usepackage{latexsym}
\usepackage{color}
\usepackage{graphicx}
\usepackage{hyperref}
\usepackage{nicefrac}
\usepackage{inputenc}
\usepackage{pstricks}
\usepackage{slashed}
\usepackage{multirow}
\usepackage{enumerate}
\usepackage{subcaption}
\usepackage[colorlinks=true,urlcolor=blue,anchorcolor=blue,citecolor=blue,filecolor=blue,linkcolor=blue,menucolor=blue,pagecolor=blue,linktocpage=true,pdfproducer=medialab,pdfa=true]{hyperref}
\def\IR{\relax{\rm I\kern-.18em R}}

\def\inv{^{\raise.0ex\hbox{${\scriptscriptstyle -}$}\kern-.05em 1}}

\begin{document}

\title{
Holographic Timelike Entanglement Across Dimensions}

\author[a]{Carlos Nunez}
\author[b]{ and ~Dibakar Roychowdhury}
\affiliation[a]{Centre for Quantum Fields and Gravity, Department of Physics, Swansea University,
Swansea SA2 8PP, United Kingdom}
\affiliation[b]{Department of Physics, Indian Institute of Technology Roorkee
Roorkee 247667, Uttarakhand,
India}

\emailAdd{c.nunez@swansea.ac.uk}
\emailAdd{dibakar.roychowdhury@ph.iitr.ac.in}
\abstract{We develop a holographic framework for computing timelike entanglement entropy (tEE) in quantum field theories, extending the Ryu–Takayanagi prescription into Lorentzian settings. 
Using three broad classes of supergravity backgrounds, we derive both exact and approximate tEE expressions for slab, spherical, and hyperbolic regions, and relate them to the central charges of the dual conformal field theories. The method is applied to infinite families of supersymmetric linear quivers in 
dimensions from $d=2$ to $d=6$, showing that Liu–Mezei and slab central charges scale universally like the holographic central charge. We then analyse gapped and confining models, including twisted compactifications and wrapped brane constructions, identifying how a mass gap modifies tEE and when approximate formulas remain accurate. In all cases, we uncover robust scaling with invariant separations and signature-dependent phase behaviour, distinguishing spacelike from timelike embeddings. Our results unify the treatment of tEE in both conformal and non-conformal theories, clarifying its role as a probe of causal structure, universal data, and non-perturbative dynamics in holography.}

\maketitle

\flushbottom


\section{Introduction and Summary}
%
%
%
%
%
%
Entanglement entropy has emerged as a fundamental probe of quantum correlations in quantum field theory (QFT) and many-body systems, offering deep insights into the structure of spacetime and quantum matter. In conformal field theories (CFTs), it provides universal information encoded in the central charges and operator content, while also serving as a bridge between statistical mechanics, condensed matter physics, and quantum gravity \cite{Kitaev:2005dm,Levin:2006zz}. The holographic correspondence \cite{Maldacena:1997re, Gubser:1998bc, Witten:1998qj} has established a powerful framework for computing entanglement entropy in strongly coupled systems, most notably through the Ryu–Takayanagi prescription \cite{Ryu:2006ef,  Ryu:2006bv, Hubeny:2007xt}, which connects geometric quantities in the bulk to information-theoretic measures in the boundary theory.

While most studies have focused on space-like entanglement, the concept of \emph{time-like} entanglement entropy (tEE) offers a new perspective into the causal structure and temporal correlations of quantum fields. In particular, CFTs in diverse dimensions offer a fertile ground for both exact and approximate calculations, especially when complemented by holographic duals of confining field theories \cite{Nakagawa:2009jk,Rabenstein:2018bri}. Such settings are sensitive to nonperturbative features and scale-dependent phenomena, making them ideal laboratories for probing the dynamics of entanglement in nontrivial geometries.

The timelike entanglement entropy is related (via Wick rotation of the usual EE) to a concept in information theory known as pseudo-entropy. Pseudo-entropy provides a framework to quantify the entanglement for a {\it transition between states}. For a careful explanation of pseudo-entropy, its calculation in field theory and holography and its relation with tEE see \cite{Nakata:2020luh, Mollabashi:2021xsd,Doi:2022iyj}. These ideas nicely relate with those presented in \cite{Milekhin:2025ycm, Gong:2025pnu}. The point is to calculate the entanglement in time as the correlation between two regions $A$ and $B$ which are separated in time. This leads to the definition of $T_{AB}$ (a generically non-hermitian operator), which is the generalisation to the time-separated case, of the density matrix--
 defined for the case in which $A$ and $B$ are separated in space. 
 
 In the case of CFT$_2$ and AdS$_3$ the connection is understood, in that case the operator $T_{AB}$ is obtained by analytic continuation of correlator of twist operators \cite{Milekhin:2025ycm, Gong:2025pnu}. In the 2d-CFT case, the tEE presents an imaginary part and a real part (that is logarithmic in the time separation). The real part comes from the usual RT-surface. The imaginary part is  understood moving to a complexified AdS-space \cite{Heller:2024whi, Heller:2025kvp}. In this vein, the papers \cite{Nunez:2025ppd, Heller:2025kvp} present a way to holographically define the tEE using the usual RT (spacelike) EE and analytically continuing across the light-cone.

There are many other motivations to study time-like entanglement entropy beyond those mentioned above. In fact, the concept of entanglement can be applied
to any arbitrary quantum system, generalising the usual density matrix to a space-time density matrix. Studying these concepts within the framework of holography seems a good idea, as the problem becomes geometrized.
On this same line, there is an interesting connection with the theory of Tensor Networks and the complexity of algorithms that represent time evolution. See for example \cite{Banuls:2009jmn, Carignano:2023xbz}

As mentioned above, at present, there is no clear consensus on the way to calculate tEE. Some  proposals, that rely both on QFT arguments and on the geometry side of the duality and guide our work, include \cite{Heller:2024whi, Xu:2024yvf, Heller:2025kvp, Nunez:2025gxq, Nunez:2025ppd, Gong:2025pnu}. This deserves further study.
Moreover, studying entanglement across timelike regions can reveal intricate links between information-theoretic measures and renormalisation group flows, including constraints from the $c$-, $a$- and $F$-theorems \cite{Zamolodchikov:1986gt, Komargodski:2011vj, Casini:2004bw, Casini:2015woa}. 

In this work, we explore timelike entanglement entropy in CFTs across dimensions, deriving exact results, developing controlled approximations, and relating our findings to the central charges of the theory. We also discuss holographic duals to field theories with a mass gap and/or confinement.
\\
The contents of this paper are organised as follows:
\begin{itemize}
    \item{In Section \ref{sectionpreliminar}, we discuss different generic aspects of EE and tEE. We present a summary of results with emphasis on the approach developed recently in \cite{Nunez:2025gxq, Nunez:2025ppd, Jokela:2025cyz}. We divide our presentation into three classes of backgrounds, that are further explored in the following sections. We provide expressions that approximate the separation and the tEE (these are useful when intensive numerical analysis is needed to evaluate the exact expressions). }
    \item{In Section \ref{sectionconformalmodels}, we specialise our formulas for holographic duals to families of CFTs in diverse dimensions (the results are presented for entanglement across slab-regions and spherical ones). We present expressions for the central charge of the CFTs, computed using the holographic description. }
    \item{Section \ref{sectionexamplescft} presents the careful study of the material in previous sections, applying it to the holographic duals to an infinite family of SCFTs (with eight Poincare supercharges). Precise expressions are presented.}
    \item{In Section \ref{sectionconfiningmodels} we extend the study of the first few sections to systems that present a mass gap (and in some cases confinement). A careful explanation of the dynamics of such systems is given.  Here, we discuss holographic duals that share the (dynamical part) of the entanglement entropy, whilst having qualitatively different UV fixed points. Some of the approximate expressions derived in previous sections become instrumental for some of these models. Otherwise, the exact expressions would require intensive numerical work, that we do not do in this work. Importantly, models with an extra scale (like the mass gap in these systems) allow for a {\it real-valued Ryu-Takayanagi surface} in the holographic calculation of  timelike entanglement entropy.}
\end{itemize}
We end with some concluding remarks and future directions in Section \ref{sectioncoclusions}.

\section{Preliminaries}\label{sectionpreliminar}
In this work, our focus is on timelike entanglement entropy. It is known that the extremal surface calculating this quantity becomes complex (the turning point of the surface does), presenting a puzzling situation. Several approaches were proposed to better understand this issue. We build on the ideas in \cite{Guo:2025pru, Heller:2024whi, Heller:2025kvp}, in particular, adopting the perspective presented in \cite{Nunez:2025ppd}. 
In this approach we consider space-time entanglement in the case of slab regions. It is not yet clear whether a similar approach can be applied to spherical regions. 

The purpose of this section is to review the formalism of space-time entanglement for slab regions.
We then specialise the formalism to the case of timelike entanglement and present expressions that can approximate both the time separation and the timelike entanglement entropy. These expressions are used in examples, presented in subsequent sections. 
We consider here  three classes of backgrounds (we refer to them as class I, class II and class III) and present expressions for the entanglement entropy for each class. Subsequent sections illustrate each class with well-known backgrounds.
\subsection{Three classes of backgrounds and space-time entanglement}
We consider three classes of supergravity backgrounds in ten dimensions (the extension to eleven-dimensional supergravity is straightforward). We write the metric and the dilaton, noting that the Neveu–Schwarz and Ramond forms complement these backgrounds but do not play a role in this work and are not written.
We work in string-frame. 
\begin{itemize}
    \item{The backgrounds of class I read,}
\end{itemize}
\begin{eqnarray}
& & ds^2_{I,st}= f(\vec{v})\text{AdS}_{d+1}+ g_{ij,(9-d)}(\vec{v})dv^i dv^j, ~~~\Phi(\vec{v}).\label{classi}\\
& &\text{AdS}_{d+1}= \frac{u^2}{l^2}(\lambda dt^2+ d\vec{x}_{d-2}^2 + dy^2)+\frac{l^2 du^2}{u^2}.\nonumber
\end{eqnarray}
We explain our conventions. The background is ten dimensional, solves the supergravity equations of motion (the extension to eleven dimensional supergravity is straightforward).
The AdS-space is $(d+1)$-dimensional. We use the (Poincare patch) coordinates as $(t,y,\vec{x}_{d-2},u)$ to describe it. The time is Euclidean (when $\lambda=1$) or Lorentzian (when $\lambda=-1$). The internal space with metric $g_{ij}(\vec{v})$ is $(9-d)$-dimensional and parametrised by the coordinates $\vec{v}$. The warp factors $f(\vec{v})$, $g_{ij}(\vec{v})$ and the dilaton $ \Phi(\vec{v})$ depend only on the internal space coordinates, otherwise they would spoil the isometries of AdS. The same holds for the NS and RR forms, which we do not include explicitly.
\begin{itemize}
\item{The backgrounds in class II read,}
\end{itemize}
\begin{eqnarray}
& & ds^2_{II,st}= f(\vec{v},u)\Big[ s_1(u)\left( \lambda dt^2+d\vec{x}^2_{d-3}+ dy^2\right) + s_2(u) du^2+ s_3(u) d\phi^2 \Big]+ \nonumber\\
& & ~~~~~~~~~~~+ g_{ij,(9-d)}(\vec{v},u)(dv^i-A^i) (dv^j-A^j), ~~~\Phi(\vec{v},u),\label{classii}\\
& &~~~~~~~~~~~~A^i= s_{4,i}(u) d\phi.\nonumber
\end{eqnarray}
Backgrounds in class II describe the generic form of solutions in Type II A/B (or M-theory) that are dual to CFTs in dimension $d$ with one direction--
 here called $\phi$ -- compactified on a circle and twisted with the R-symmetry to preserve some SUSY. This is realised in holography with the $\phi$-direction being fibered (by virtue of the one forms $A^i$), over the internal $(9-d)$ dimensional space. These backgrounds, dual to CFTs that flow to a gapped (usually confining) IR QFT in $(d-1)$-dimensions are exemplified by  \cite{Anabalon:2021tua, Anabalon:2022aig,Anabalon:2024che, Anabalon:2024qhf, Nunez:2023nnl, Nunez:2023xgl, Chatzis:2024kdu, Chatzis:2024top, Chatzis:2025dnu, Fatemiabhari:2024aua, Kumar:2024pcz, Macpherson:2024qfi}.
\begin{itemize}
    \item{Finally, the class III backgrounds read,} 
\end{itemize}
\begin{eqnarray}
& & ds^2_{III, st}=\hat{h}(u)^{-\frac{1}{2}}\left( \lambda dt^2+ dy^2+d\vec{x}_{d-2}^2\right) + \hat{h}(u)^{\frac{1}{2}}e^{2k(u)} du^2+ g_{ij,(9-d)}(u,\vec{v})dv^i dv^j, ~~\Phi(u).\label{classiii}    
\end{eqnarray}
These backgrounds are simpler, describing holographic duals to field theories (characteristically confining ones) that are obtained on wrapped brane systems or on intersections of branes and fractional branes. Examples include the backgrounds in references \cite{Maldacena:2000yy, Klebanov:2000hb, Maldacena:2001pb, Edelstein:2001pu, Casero:2006pt, Casero:2007jj, Hoyos-Badajoz:2008znk, Benini:2006hh, Butti:2004pk, Bobev:2018eer, Petrini:2018pjk}.
\\
\\
In what follows we apply the formalism developed in \cite{Nunez:2025ppd} to calculate the entanglement entropy of a space-time slab for the three classes of backgrounds above. It would be interesting to extend this to the case of Lifshitz-like and other non-relativistic backgrounds \cite{Afrasiar:2024ldn}. In order to do this we choose an eight surface (it would be a nine-surface for M-theory backgrounds). For the three classes of backgrounds, the eight-surface is parametrised by
\begin{eqnarray}
& & \Sigma_{8,I}=\Sigma_{8,III}=[u, \vec{x}_{d-2},\vec{v}_{9-d}],~~
\Sigma_{8,II}=[u, \vec{x}_{d-3},\phi,\vec{v}_{9-d}],\nonumber\\
& &~\text{with}~~t(u), y(u).\label{surfaces}
\end{eqnarray}

Then we calculate the induced metric on each surface,
\begin{eqnarray}
& & ds^2_{\Sigma_{8,I}}= f(\vec{v})\left(  du^2\left[\frac{l^2}{u^2}+ \frac{u^2}{l^2}(\lambda t'^2+y'^2) \right]+ \frac{u^2}{l^2} d\vec{x}^2_{d-2} \right)+ g_{ij,(9-d)}(\vec{v})dv^i dv^j, \label{metricsigma8s}\\
& & ds^2_{\Sigma_{8,II}}= f(\vec{v}, u)\Big(  du^2\left[s_2(u)+ s_1(u)(\lambda t'^2+y'^2) \right]+ s_1(u)  d\vec{x}^2_{d-3} + s_3(u) d\phi^2\Big)+ g_{ij,(9-d)}(\vec{v},u)(dv^i-A^i)( dv^j-A^j), \nonumber\\
& & ds^2_{\Sigma_{8,III}}=   du^2\Big[\hat{h}(u)^{\frac{1}{2}} e^{2k}+ \hat{h}(u)^{-\frac{1}{2}}(\lambda t'^2+y'^2) \Big]+ \hat{h}(u)^{-\frac{1}{2}} d\vec{x}^2_{d-2} + g_{ij,(9-d)}(\vec{v},u)dv^i dv^j. \nonumber
\end{eqnarray}
After this, one calculates the U-duality invariant quantity (recall that we work in string-frame, in Einstein frame, the dilaton factor below is absent),
\begin{eqnarray}
& & 4 G_{10} S_{EE}=\int_{\Sigma_8} \sqrt{e^{-4\Phi} \det[g_{ind,\Sigma_8}]}.    
\end{eqnarray}
Here $G_{10}=8\pi^6 g_s^2\alpha'^4$ is the ten dimensional Newton constant. For the backgrounds of class I we find,
\begin{eqnarray}
& & e^{-4\Phi}\det[g_{\Sigma_{8,I}}]=e^{-4\Phi(\vec{v})} f(\vec{v})^{d-1}\det[g_{9-d}] \Big[ \left(\frac{u}{l}\right)^{2(d-3)} + \left( \frac{u}{l}\right)^{2(d-1)} (y'^2+\lambda t'^2) \Big],\nonumber\\
& &   4 G_{10} S_{EE,I}={\cal N}_I\int du \sqrt{F^2(u)(y'^2+\lambda t'^2)+ G^2(u)},~~\text{where we defined}\label{SEEI}\\
  & & F(u)=\left(\frac{u}{l}\right)^{(d-1)},~~G(u)=\left(\frac{u}{l}\right)^{(d-3)},~{\cal N}_I= \int d^{9-d}v ~d^{d-2}x\sqrt{e^{-4\Phi(\vec{v})} f(\vec{v})^{d-1}\det[g_{9-d}]}.\nonumber
\end{eqnarray}
A similar result is obtained for the backgrounds in classes II and III.
In fact, 
\begin{eqnarray}
& &e^{-4\Phi}\det[g_{\Sigma_{8,II}}]=e^{-4\Phi(\vec{v},u)} f(\vec{v},u)^{d-1}~\det[g_{9-d}(\vec{v},u)]~s_3(u)~s_1(u)^{d-3}\left[ s_1(u)(\lambda t'^2+y'^2) + s_2(u)\right],\nonumber\\
& &e^{-4\Phi}\det[g_{\Sigma_{8,III}}]=e^{-4\Phi(u)} \det[g_{9-d}(\vec{v},u)] ~\hat{h}(u)^{\frac{(1-d)}{2}}~\left[ \lambda t'^2+y'^2 + e^{2k}~\hat{h}(u)\right].\label{class23-bis}
\end{eqnarray}

%
As we discuss in Section \ref{sectionconfiningmodels}, an interesting conspiracy occurs and the integral over the internal space parametrised by the $\vec{v}$-coordinates, can be performed explicitly, leading for the three classes of backgrounds to a one dimensional action of the form,
\begin{eqnarray}
& & 4 G_{10} S_{EE}={{\cal N}} \int_{u_0}^\infty~ du \sqrt{F^2(u) (y'^2 +\lambda t'^2) + G^2(u)}.\label{SEEwork}    
\end{eqnarray}

Without specifying the functional form of $\Phi(\vec{v},u), f(\vec{v},u)$ and $g_{ij,(9-d)}(\vec{v},u)$ it is not possible to be more precise about the functions $F(u)$ and $G(u)$. 
The functions $F(u),G(u)$ and the constant ${\cal N}$ do depend on the particular supergravity solution in question. We explore various examples in Sections \ref{sectionconformalmodels}, \ref{sectionexamplescft} and \ref{sectionconfiningmodels}. In what follows, we focus on actions of the form (\ref{SEEwork}). In generic terms we detail the treatment in the papers \cite{Nunez:2025gxq, Nunez:2025ppd}.
\\
In what follows we work with the action in eq.(\ref{SEEwork}).
The quantity $\frac{{\cal N}}{4 G_{10}}$ is related to the central charge of the dual UV-CFT (and dependent on the class of backgrounds we discuss). The  parameter $u_0$ is the turning point of the eight-surface, the lowest value of the coordinate $u$ reached by $\Sigma_8$. The equations of motion of the variables $t(u), y(u)$ are,
\begin{eqnarray}
& &    \frac{F^2 t'}{\sqrt{G^2+F^2 (\lambda t'^2 +y'^2)}}=\lambda c_t,~~~~
    \frac{F^2 y'}{\sqrt{G^2+F^2 (\lambda t'^2 +y'^2)}}= c_y.\nonumber
\end{eqnarray}
Here $c_y$ and $c_t$ are constants of motion.
The solution to these equations read
\begin{eqnarray}
 &   &t'^2(u)=\frac{c^2_t G^2(u)}{F^2(u)(F^2(u)-F^2(u_0))},~~~~y'^2 (u)=\frac{c^2_y G^2(u)}{F^2(u)(F^2(u)-F^2(u_0))}.\label{eq3}
    \end{eqnarray}
We denote the turning point (this is the point at which both $t'$ and $y'$ diverge)
\begin{align}
    F^2(u_0)=c^2_y+\lambda c^2_t.
\end{align}
We  express  the lengths of the subsystems (in the coordinates $t$ and $y$ respectively)
\begin{eqnarray}
& &     T=2 c_t\int_{u_0}^{\infty}du \frac{G(u)}{F(u)\sqrt{F^2(u)-F^2(u_0)}},~~~ Y=2 c_y\int_{u_0}^{\infty}du \frac{G(u)}{F(u)\sqrt{F^2(u)-F^2(u_0)}}.\label{separationsTY}
\end{eqnarray}
The entanglement entropy after regularisation is \cite{Nunez:2025ppd},
\begin{align}
  \frac{4G_{10}}{\cal N}  S_{tEE}=2\int_{u_0}^{\infty}du \frac{G(u)F(u)}{\sqrt{F^2(u)-F^2(u_0)}}-2\int_{u_\ast}^\infty G(u)du.\label{regulatedEE}
\end{align}
In what follows we mostly focus on the Lorentzian signature case, hence choose $\lambda=-1$. The turning point is given by $ F^2(u_0)=c^2_y - c^2_t$. On the other hand, the point $u_*$ denotes the end of the space, being $u_*=0$ for AdS-Poincare coordinates. For generic metrics like those in class II or III the point $u_*$ depends on the place where some cycle (like the $\phi$-coordinate) shrinks. 

\begin{figure}
    \centering
    \includegraphics[width=0.4\linewidth]{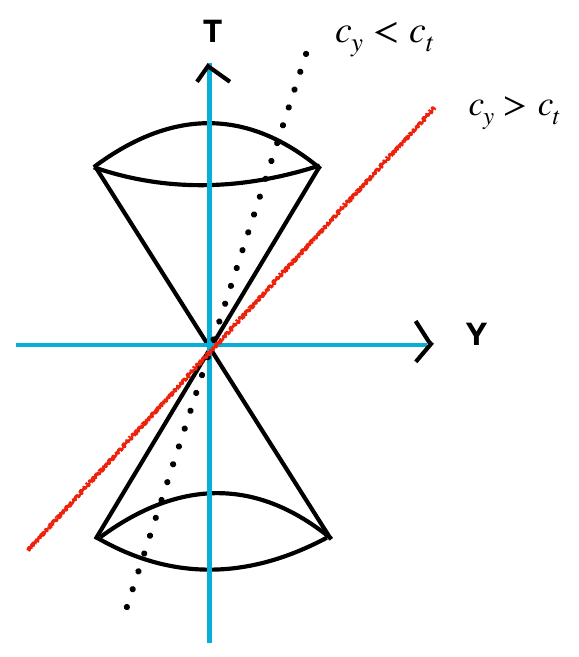}
    \caption{Light cone structure that clearly distinguishes between spacelike and timelike separated events. The red line ($c_y>c_t$) corresponds to spacelike separated events (and hence a Type I or usual RT like extremal surface). On the other hand, the dotted line ($c_y<c_t$) corresponds to a timelike separated events that corresponds to a Type II (or complex) extremal surface. Clearly, $c_y=0$ corresponds to pure timelike separated events that are along the time ($T$) axis of the diagram.}
    \label{figLC}
\end{figure}

From the above discussion, it is generic that the turning point in Lorentzian cases $F^2(u_0)=c_y^2-c_t^2$ is real for $c_y>c_t$. In this case we have a spacelike separated slab. For $c_t>c_y$ we have a time-like separated slab and the turning point is generically imaginary. In \cite{Nunez:2025ppd} the two types of embeddings were named Type I and Type II respectively. {We anticipate here that for the case of dual QFTs with a scale (such as confining or gapped theories), real embeddings (Type I) with real turning points $u_0$, even for the case $c_y<c_t$. A more detailed discussion is given in Section \ref{sectionconfiningmodels}. }

The idea is to write both the $S_{tEE}(u_0)$ in eq.(\ref{regulatedEE}) and the space-time interval $\Delta^2(u_0)=Y^2(u_0)-T^2(u_0)$ from eq.(\ref{separationsTY}) and then parametrically write $S_{tEE}(\Delta)$. For $\Delta^2>0$ we have  Type I embeddings, for $\Delta^2<0$ Type II embeddings. The lightlike case is analysed as a limiting procedure  in \cite{Nunez:2025ppd}. In that paper, a way is suggested to interpolate between the two types of embeddings via analytic continuation to avoid the divergent result on the light-cone, something similar to the proposal of \cite{Heller:2025kvp}. In other words, we start with a real valued embedding (Type I) with $c_y>c_t$ and increase $c_t$. The case $\Delta^2=0$ is avoided by analytic continuation. After that we continue to increase $c_t$ and we have a Type II, complex embedding. 

Below, we mostly focus on the purely time-like situation, that is when $c_y=0$, leading to Type II embeddings. Of course the character of the embedding depends on the precise functional form of $F^2(u)$. In cases in which there is a length scale in the field theory (like the examples studied in Section \ref{sectionconfiningmodels}), the system can have Type I embeddings even if $c_t>c_y$.
\subsection{The case of time-like Entanglement}\label{tEEsection}
We now specialise our expressions in eqs.(\ref{SEEwork})-(\ref{regulatedEE}) to the case of purely time-like entanglement. In this case the integration constant $c_y=0$, in other words, the coordinate $y$ is fixed. When $y'=0$, the action in eq.(\ref{SEEwork}) has the same structure as the action obtained when probing the background with a fundamental string (to calculate the quark-antiquark pair separation and energy). In the case of space-like EE (with $t'=c_t=0$), this formal similarity was exploited in \cite{Kol:2014nqa} to write expressions that well-approximate the separation of the slab. In the same vein, the paper \cite{Nunez:2025gxq} made use of the formal similarity to write approximate expressions for the time-like separation and the time-like EE. Below, we summarise these expressions, as they become useful in subsequent sections.

Consider an action of the form in eq.(\ref{SEEwork}), for the special case of fixing the value of the $t$-coordinate, $t'=0$. In \cite{Kol:2014nqa, Faedo:2013ota} it was proposed (without a proof)  that the expression $Y_{app}(u_0)=\pi \frac{G(u_0)}{F'(u_0)} $ {\it approximates} the separation in $Y$, otherwise given by the integral in eq.(\ref{separationsTY}), in terms of the turning point $u_0$. Along this line, for the case $y'=0$, the {\it approximate} time separation $T_{app}(u_0)$ is \cite{Nunez:2025gxq},
\begin{equation}
 T_{app}(u_0)=\pi\frac{G(u_0)}{F'(u_0)}.\label{Tapp}   
\end{equation}
In cases for which the evaluation of the integrals in eq.(\ref{separationsTY}) is complicated, one can resort to the 'experimentally motivated' expression in eq.(\ref{Tapp}). It would be nice to provide a proof of eq.(\ref{Tapp}).
\\
Similarly,  in the context of Wilson loops \cite{Nunez:2009da, Faedo:2013ota}, it was shown  the {\it exact} expression for the functional $S(u_0)$ --like the one in eq.(\ref{regulatedEE}), satisfies
\begin{equation}
\frac{dS(u_0)}{dY(u_0)}= F(u_0).\label{notso}    
\end{equation}
Although obtained in the study of Wilson loops, the proof given in \cite{Nunez:2009da, Faedo:2013ota} also applies to the case of entanglement entropy, due to the formal similarity between the actions (for the F1 string and for the eight-surface).

Using eqs.(\ref{Tapp}),(\ref{notso}) in the case $c_y=0$
we follow \cite{Nunez:2025gxq} and find,
\begin{eqnarray}
& &    \frac{dS(u_0)}{dT(u_0)}\approx \frac{dS_{approx}(u_0)}{dT_{app}(u_0)}\approx F(u_0),~~\text{integrating we find,}\nonumber\\
& &S_{approx}(u_0)= \int F(u_0) dT_{app}(u_0)= \pi\int F(u_0) \frac{d}{du_0}\left(\frac{G(u_0)}{F'(u_0)}\right) du_0 +~\text{constant}.\label{Sapp}
\end{eqnarray}
When studying holographic duals to non-conformal field theories, it is usually the case  that the integrals giving the exact expressions for the separations and the EE, eqs.(\ref{separationsTY}),(\ref{regulatedEE}) are not easy to evaluate and one needs to resort to (typically demanding) numerical work. It is in those cases that the {\it approximate} expressions in eqs.(\ref{Tapp}) and (\ref{Sapp}) are particularly useful. 

The analysis in this preliminary section is based on slab-regions. In the case of conformal field theories, the integrals can be performed explicitly not only in the case of slabs, but also for spherical or hyperbolic regions. We study these expressions below, in Section \ref{sectionconformalmodels}.

\section{(Timelike) Entanglement for generic CFT$_d$}\label{sectionconformalmodels}
In this section, we discuss the general formulae  above, for conformal field theories in $d$ space-time directions. We work with backgrounds in class I, eq.(\ref{classi}) containing AdS$_{d+1}$ subspaces. We begin with the calculation of the entanglement entropy on a slab region. We write exact expressions for the interval separation and the EE in terms of the turning point in the bulk. We then provide a useful expression for the EE in terms of the invariant separation (the interval $\Delta^2=Y^2-T^2$), expressing everything in terms of field theory quantities. 

After this, we discuss the calculation of the entanglement entropy in spherical or hyperbolic regions, quoting exact expressions derived in \cite{Doi:2023zaf,Nunez:2025gxq}. We present  expressions for the central charge of these systems, derived from the entanglement entropy.

\subsection{Entanglement entropy for holographic CFTs in dimension $d$ }
We consider class I backgrounds, which are dual to conformal field theories in $d$ spacetime dimensions.
The background and dilaton read,
\begin{eqnarray}
& & ds^2_{I,st}= f(\vec{v})\text{AdS}_{d+1}+ g_{ij,(9-d)}(\vec{v})dv^i dv^j, ~~~\Phi(\vec{v}).\label{classibis}
\end{eqnarray}
The AdS-subspace is written as,
\begin{eqnarray}
& &\text{AdS}_{d+1}= \frac{u^2}{l^2}(\lambda dt^2+ d\vec{x}_{d-2}^2 + dy^2)+\frac{l^2 du^2}{u^2},\label{adsp}\\
& & \text{or},\nonumber\\
& & {AdS_{d+1}}=u^2(\lambda dt^2+t^2 d\Omega^{(\lambda)}_{d-2}+dy^2)+\frac{du^2}{u^2}.\label{adscurved}
\end{eqnarray}
The first expression is used for slab regions, and the second for spherical ($\lambda=+1$, Euclidean signature) or hyperbolic regions ($\lambda=-1$, Lorentzian signature). {For Euclidean time ($\lambda=+1$), the subspace $d\Omega_{d-2}^{(\lambda=+1)}$ in eq.(\ref{adscurved}) is a sphere. The resulting metric is that of Euclidean AdS$_{d+1}$. In contrast, for $\lambda=-1$ (Lorentzian signature) the space $d\Omega_{d-2}^{(\lambda=-1)}$ is a hyperbolic plane. The resulting metric is that of Lorentzian AdS$_{d+1}$. }
\subsubsection{Slab-regions}
We focus on slab regions (with Lorentzian signature). The induced metric is that in the first line of  eq.(\ref{metricsigma8s}) and the entanglement entropy is given in eq.(\ref{SEEI}), for $\lambda=-1$. We can explicitly evaluate  eqs.(\ref{separationsTY})-(\ref{regulatedEE}) using eq.(\ref{SEEI}) for the functions $F(u), G(u)$
\begin{eqnarray}
& &  T=2 c_t ~l^{d+1} \int_{u_0}^{\infty}\frac{du}{u^2}\frac{1}{\sqrt{u^{2(d-1)}-u_0^{2(d-1)}}},~~~~
Y=2 c_y~ l^{d+1} \int_{u_0}^{\infty}\frac{du}{u^2}\frac{1}{\sqrt{u^{2(d-1)}-u_0^{2(d-1)}}}.\label{Tads}
\end{eqnarray}
For the EE we find,
\begin{eqnarray}
\frac{4 G_{10}}{{\cal N}}S_{EE}&=&    \frac{2}{l^{d-3}}\Bigg[\int_{u_0}^\infty du \frac{u^{2d-4}}{\sqrt{u^{2d-2} - u_0^{2d-2} }} - \int_0^\infty u^{d-3}~ du \Bigg].\label{SEEAdS}
\end{eqnarray}
The constant ${\cal N}={\cal N}_I$ in eq.(\ref{SEEI}). The two possible embeddings are characterised by the value of
\begin{equation}
u_0= \left( c_y^2-c_t^2\right)^{\frac{1}{2(d-1)}}.\label{u0gen}
\end{equation}
The  Type I embeddings have $(c_y^2-c_t^2)>0$. The turning point occurs at a real value of the coordinate $u=u_0$ in eq.(\ref{u0gen}). In contrast, for Type II embeddings (when $c_t^2>c_y^2$) we have
\begin{equation}
u_0= e^{\frac{i\pi}{2(d-1)}}\left[\sqrt{|c_y^2-c_t^2|}\right]^{\frac{1}{d-1}}=  \left( i \tilde{u}_0\right)^{\frac{1}{d-1}}.   
\end{equation}
The turning point is at a complex value of the $u$-coordinate, note that $\tilde{u}_0 =\sqrt{|c_y^2-c_t^2|}$ is real.
%
%
%
We change to the variable $r=\frac{u_0}{u}$. The expressions for the $T$-separation and $Y$-separation are, 
\begin{eqnarray}
& &  T=\frac{2 c_t~ ~l^{d+1}}{u^d_0} I_1,~~~Y= \frac{2 c_y~~ l^{d+1}}{u_0^d} I_1,~\text{where}\nonumber\\
& &  I_1=\int_0^1 dr \frac{r^{d-1}}{\sqrt{1-r^{2(d-1)}}}=\frac{\sqrt{\pi } \Gamma \left(\frac{d}{2 d-2}\right)}{\Gamma \left(\frac{1}{2 (d-1)}\right)}.\label{YT-ads}
\end{eqnarray}
For the EE in eq.(\ref{SEEAdS}) we find,
\begin{eqnarray}
& & \frac{2 G_{10} ~l^{d-3}}{{\cal N}~ u_0^{d-2}}S_{EE}= I_2-I_3, ~~~\text{where}\label{SEEAdS2} \\  
& & I_2= \int_\epsilon^1 \frac{dr}{r^{d-1}\sqrt{1- r^{2d-2}}}
= -\frac{1}{(d-2)r^{d-2}} \, _2F_1\left(\frac{1}{2},\frac{2-d}{2 d-2};\frac{d}{2 (d-1)};r^{2 d-2}\right) \Bigg|_\epsilon^1,\nonumber\\
& & I_3= \int_{\epsilon}^\infty \frac{dr}{r^{d-1}}=\frac{1}{(d-2)~\epsilon^{d-2}}.\nonumber
\end{eqnarray}
We introduced the small parameter $\epsilon\to 0$ to UV-regulate the quantities $I_2, I_3$. One can check that the divergence (for $\epsilon\to 0$) in $I_2$ is precisely cancelled by the divergence  in $I_3$. This is the logic of the UV-regulation  in eq.(\ref{regulatedEE}). The result is
\begin{eqnarray}
 \frac{2 G_{N,10} ~l^{d-3} }{{\cal N}~ u_0^{d-2}}\!S_{EE}
 =\frac{1}{(2-d)} ~{}_2F_1\left(\frac{1}{2}, \frac{2-d}{2d-2};\frac{d}{2d-2};1 \right).  \label{SEEI2I3} 
\end{eqnarray}

{Before proceeding further, it is worth mentioning that for pure time-like separation ($c_y=0$), we have two branches of solutions ($T_{\pm}$) for the time-like slab as also observed in \cite{Heller:2024whi}. In our calculation, this can be seen from \eqref{u0gen}, which yields $c_t=\pm i u_0^{d-1}$. When substituted back into \eqref{YT-ads}, this produces a lower ($T_-$) and an upper ($T_+$) branch of solutions that meet smoothly at the turning point $u_0$.}

We write the entanglement entropy in terms of the physical quantities in the field theory, namely the separations $Y$ and $T$. To do this we find the integration constants $(c_y,c_t)$ from eq.(\ref{YT-ads}) and put this together with eq.(\ref{u0gen}) to obtain,
\begin{eqnarray}
& & c_t= \frac{T}{2 l^{d+1} I_1} u_0^d,~~ c_y= \frac{Y}{2 l^{d+1} I_1} u_0^d,~~~ u_0=\frac{2 l^{d+1} I_1}{\sqrt{Y^2-T^2}}.\label{u0generico}   
\end{eqnarray}
Using eq.(\ref{SEEI2I3}) gives,
\begin{eqnarray}
& &S_{EE}= \frac{{\cal N} (I_2-I_3) 2^{d-3} l^{(d-1)^2} I_1^{d-2}}{G_{10}} \times \frac{1}{\left( Y^2-T^2\right)^{\frac{(d-2)}{2}}}   .\label{SEEAdSfinal}\\
& & \text{and}\nonumber\\
& &S_{EE}= \frac{{\cal N} (I_2-I_3) 2^{d-3} l^{(d-1)^2} I_1^{d-2}}{G_{10}} \times \frac{e^{-i\pi\frac{(d-2)}{2}}}{ |Y^2-T^2|^{\frac{(d-2)}{2}}}   .\label{SEEAdSfinal2}
\end{eqnarray}
 If the interval  $\Delta^2= Y^2- T^2$ is positive we are considering surfaces with a Type I embedding (as $c_y>c_t$). We find a real result  in eq.(\ref{SEEAdSfinal}). For the case of negative  $\Delta^2$ (that corresponds to $c_y < c_t$), we are considering surfaces with a Type II embedding, in which case the result is that in eq.(\ref{SEEAdSfinal2}).
 
In the case $c_t=0$, we are in the pure Ryu-Takayanagi case
with $S_{EE}\propto \frac{1}{Y^{d-2}}$. On the other hand, for $c_y=0$, we are in the case of a Type II embedding and we find $S_{EE}\propto \frac{e^{-i\pi\frac{(d-2)}{2}}}{|T|^{d-2}} $, reproducing the result in \cite{Doi:2023zaf, Nunez:2025gxq}. Note that the result of eq.(\ref{SEEAdSfinal2}) is imaginary for odd $d$. The expressions in eqs.(\ref{SEEAdSfinal}) and (\ref{SEEAdSfinal2}) precisely match those in \cite{Nunez:2025gxq} and \cite{Doi:2023zaf} after setting $Y=0$.

In the case of purely time-like EE ($c_y=0$), we can compare the exact expressions in eqs.(\ref{YT-ads})-(\ref{SEEAdSfinal2}) with the approximate expressions in eqs.(\ref{Tapp}),(\ref{Sapp}). Using that $F(u)= \sqrt{\lambda} \left(\frac{u}{l}\right)^{d-1}$ and $G(u)=\left(\frac{u}{l}\right)^{d-3}$, 
we find 
\begin{align}
\label{e20}
  &T_{app}= \pi \frac{G(u_0)}{F'(u_0)}= \frac{\pi}{(d-1)\sqrt{\lambda} ~~u_0},\\
  &S_{EE,approx}= \int^{u_0} dz ~F(z)T_{app}'(z)= -\frac{\pi}{(d^2-3d+2)} u_0^{d-2}.
  \label{e21}
\end{align}
 It is worth mentioning that the approximate EE \eqref{e21} is defined up to an integration constant.
We have kept the parameter $\lambda=\pm 1$. The Lorentzian case ($\lambda=-1$) shows an imaginary value in the approximate separation in terms of the turning point $u_0$, matching the result  obtained  from eq.(\ref{u0generico}) in the limit $Y\to 0$.
Finally, we combine eqs.\eqref{e20} and \eqref{e21} to obtain 
   \begin{align}
   \label{e22}
       S_{EE,approx}=-\frac{\pi^{d-1}}{\sqrt{\lambda^{d-2}}(d-1)^{d-2}(d^2-3d+2)}\frac{1}{|T_{app}|^{d-2}},
   \end{align}
which coincides with eqs.(\ref{SEEAdSfinal})-(\ref{SEEAdSfinal2}), as far as dependencies on $\lambda$ and $|T|$ are concerned.
Let us now study the EE in the case of hyperbolic or spherical regions
\subsubsection{Hyperbolic or spherical regions}
We  now consider the special case in which the entangling region is either a sphere or a hyperboloid. The time direction is Euclidean (for the case of the sphere) or Lorentzian (for the hyperbolic case). {{In fact, this is a mapping of the spherical entangling region of \cite{ Heller:2025kvp} to a hyperbolic plane that expands in real time.}} To accommodate these, we retain the parameter $\lambda=\pm 1$ in our expressions. We use AdS$_{d+1}$ written in the form of eq.(\ref{adscurved}) and also take the coordinate $y$ fixed (equivalently, $c_y=0$). 

The metric of the corresponding eight-manifold is given by
\begin{align}
    ds^2_8|_{{\Sigma}_8}=f(\vec{v})\left[1+\lambda u^4 t'^2(u)\right]\frac{du^2}{u^2}+f(\vec{v})u^2 t^2 d \Omega^{(\lambda)}_{d-2}+g_{ij}(\vec{v})dv^i dv^j. 
\end{align}
The EE is given by
\begin{align}
\label{e24}
    S^{(\lambda)}_{EE}[{\Sigma}_8]=\frac{1}{4G_{10}}\int d^8 x \sqrt{e^{-4 \Phi}\det g_8}= \frac{{\hat{\mathcal{N}}}}{4G_{10}}\int du u^{d-3}t^{d-2}\sqrt{1+\lambda u^4 t'^2(u)}.
\end{align}
We defined the constant $\hat{{\cal N}}$ as
\begin{align}
   \hat{\mathcal{{N}}}=\text{Vol}(\Omega^{(\lambda)}_{d-2})\int d^{9-d}v \sqrt{e^{-4\Phi}\det g_{ij}}f^{\frac{d-1}{2}}(\vec{v}).
\end{align}

The equation of motion that  follows from eq.\eqref{e24} and its solution are,
\begin{eqnarray}
\label{e26}
  & &   \lambda  u^3 t(u) \left((d-1) \lambda  u^4 t'(u)^3+(d+1) t'(u)+u t''(u)\right)-(d-2) \left(\lambda  u^4 t'(u)^2+1\right)=0.\\
  & &  t(u)= \frac{\sqrt{R^2 u^2-\lambda }}{u}, ~~~\text{with}~~\lambda=\pm 1.
\end{eqnarray}
The boundary condition used for this solution is $t(u\to\infty)=R$, the radius of the ball-region.
\\
Substituting back into \eqref{e24}  and after the change of variables $u=\frac{\sqrt{\lambda}x}{R}$, yields
\begin{align}
\label{e30}
    \frac{4G_{10}}{{\hat{\mathcal{N}}}\sqrt{\lambda^{d-2}}}S^{(\lambda)}_{EE}[{\Sigma}_8]=\int_1^{\frac{R}{\sqrt{\lambda}\epsilon}}dx (x^2-1)^{\frac{d-3}{2}}.
\end{align}
For odd $d$, the integral \eqref{e30} results in \cite{Jokela:2025cyz}
\begin{align}
\label{e31}
   \frac{4G_{10}}{\hat{{\mathcal{N}}}\sqrt{\lambda^{d-2}}}S^{(\lambda)}_{EE}[{\Sigma}_8 ]=  \sum_{j=0}^{[\frac{d-3}{2}]}\frac{\Big( \frac{3-d}{2}\Big)_j}{j! (d-2j -2)}\Big(\frac{R}{\sqrt{\lambda}\epsilon} \Big)^{d-2j-2}-(-1)^{\frac{d+1}{2}}\frac{\sqrt{\pi}
    \Gamma\left(\frac{(d-1)}{2}\right)}{2\Gamma \left(\frac{d}{2} \right)}
\end{align}
where $\Big( \frac{3-d}{2}\Big)_j=\frac{\Gamma \Big(\frac{3-d}{2}+j\Big)}{\Gamma \Big(\frac{3-d}{2}\Big)}$ is the Pochhammer symbol.
For even $d$, the integral in eq.\eqref{e30} yields
\begin{align}
\label{e32}
    \frac{4G_{10}}{\hat{\mathcal{N}}\sqrt{\lambda^{d-2}}}S^{(\lambda)}_{EE}[\Sigma_8]&=\sum_{j=0}^{[\frac{d-3}{2}]}\frac{\Big( \frac{3-d}{2}\Big)_j}{j! (d-2j -2)}\Big(\frac{R}{\sqrt{\lambda}\epsilon} \Big)^{d-2j-2}\nonumber\\
    &-\frac{\Gamma\left(\frac{d-1}{2} \right)}{\Gamma \left(\frac{d}{2} \right)}\frac{(-1)^{d/2}}{\sqrt{\pi}}\Big(\log (\frac{2R}{\epsilon \sqrt{\lambda}}) +\frac{1}{2}\mathcal{H}_{\frac{d-2}{2}}\Big)
\end{align}
where $\mathcal{H}_{n}=1+\frac{1}{2}+\frac{1}{3}+\cdots+\frac{1}{n}$ is a Harmonic number. These results coincide with those in \cite{Nunez:2025gxq}, and after some manipulations, they can be shown to be equivalent to those in \cite{Doi:2023zaf}.
\\
Let us now discuss the central charge of these CFTs, obtained from the results for the EE we found above.
\subsubsection{Central charge}
Given the EE, one can calculate the Liu-Mezei central charge \cite{Liu:2012eea} for spherical or hyperboloid-entangling surfaces. This relates the prefactor ${\mathcal{N}}$ to the central charge of the dual superconformal field theory. In fact, for odd dimension $d$, the Liu-Mezei formula is expressed as
\begin{align}
\label{e33}
    (d-2)!! c_{LM,odd}=(R\partial_R -1)\cdots (R\partial_R -d+2)S^{(\lambda)}_{EE}[{\Sigma}_8].
\end{align}
For even dimension $d$, the Liu-Mezei central charge charge reads,
\begin{align}
\label{e34}
    (d-2)!! c_{LM,even}=R\partial_R \cdots (R\partial_R -d+2)S^{(\lambda)}_{EE}[{\Sigma}_8].
\end{align}
When using the purely time-like entanglement entropy, we take the absolute value of the expressions in eqs.\eqref{e33} and \eqref{e34} to define a meaningful quantity. 
\\
\\
In the same vein, one can use the EE of slab regions to compute the central charge of the dual conformal theory. This is expressed \cite{Jokela:2025cyz} as,
\begin{align}
    c_{slab}= \kappa \frac{T^{d-2}}{L^{d-2}}T\partial_T  S^{(\lambda)}_{EE}[\Sigma_8]
\end{align}
where $\kappa$ is a constant of proportionality. 
This yields a generic expression for the central charge of a CFT in dimension $d$ (even or odd), which reads
\begin{align}
\label{e36}
  c_{slab}=\frac{\mathcal{N}}{4G_{10}} \frac{\kappa}{L^{d-2}}\Big(\frac{ 2\sqrt{\pi} \Gamma \left(\frac{d}{2 d-2}\right)}{\Gamma \left(\frac{1}{2 (d-1)}\right)} \Big)^{d-1}\frac{1}{\sqrt{\lambda^{d-2}}}.  
\end{align}
Below, we elaborate on  the above results using explicit examples of different backgrounds with an $AdS_{d+1}$ subspace that preserve some amount of supersymmetry. 
In fact, we apply these results to different infinite families of conformal field theories in diverse dimensions. 
Our objective is to examine how these expressions appear in different dimensions and to relate the CFT central charge to the geometry.
\section{Holographic SCFTs in diverse dimensions}\label{sectionexamplescft}
In this section, we study infinite families of supersymmetric and conformal field theories (SCFTs) in various dimensions. The field theories considered are of the linear quiver type. A localisation plus matrix model treatment can be done for the case in which the number of Poincare SUSYs is eight, see for example \cite{Nunez:2023loo, Akhond:2022awd, Akhond:2022oaf}. 

\begin{figure}
\begin{center}
	\begin{tikzpicture}
	\node (1) at (-4,0) [circle,draw,thick,minimum size=1.4cm] {N$_1$};
	\node (2) at (-2,0) [circle,draw,thick,minimum size=1.4cm] {N$_2$};
	\node (3) at (0,0)  {$\dots$};
	\node (5) at (4,0) [circle,draw,thick,minimum size=1.4cm] {N$_{P}$};
	\node (4) at (2,0) [circle,draw,thick,minimum size=1.4cm] {N$_{P-1}$};
	\draw[thick] (1) -- (2) -- (3) -- (4) -- (5);
	\node (1b) at (-4,-2) [rectangle,draw,thick,minimum size=1.2cm] {F$_1$};
	\node (2b) at (-2,-2) [rectangle,draw,thick,minimum size=1.2cm] {F$_2$};
	\node (3b) at (0,0)  {$\dots$};
	\node (5b) at (4,-2) [rectangle,draw,thick,minimum size=1.2cm] {F$_P$};
	\node (4b) at (2,-2) [rectangle,draw,thick,minimum size=1.2cm] {F$_{P-1}$};
	\draw[thick] (1) -- (1b);
	\draw[thick] (2) -- (2b);
	\draw[thick] (4) -- (4b);
	\draw[thick] (5) -- (5b);
	\end{tikzpicture}
\end{center}
    \caption{A linear quiver. The balancing condition implies $F_i = 2 N_i - N_{i-1}-N_{i+1}$}
    \label{fig:quiver}
\end{figure}
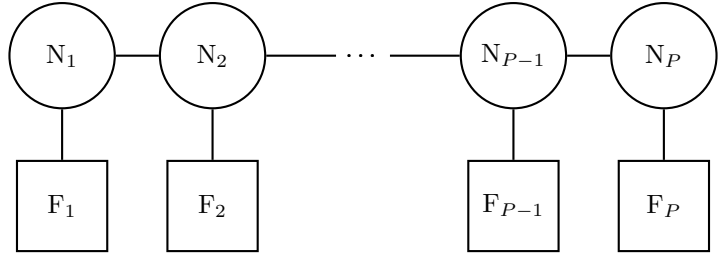

An alternative perspective for the field theories arises from  Hanany-Witten set-ups \cite{Hanany:1996ie}. These  consist of $(P+1)$ Neveu-Schwarz five branes. Between two consecutive five branes $N_i$ D$_q$ branes extend, playing the role of colour (gauged) nodes. Also $F_i$ D$_{q+2}$ branes are localised in between
consecutive NS-five branes and play the role of flavour (global) nodes. In field theoretical terms, we have linear quivers like those in Figure \ref{fig:quiver}.
\\
We consider the case of {\it balanced quivers}. This means that the condition 
\begin{equation}
F_i = 2 N_i - N_{i-1}-N_{i+1},\label{balanced} \end{equation}
is satisfied for all nodes. This condition in eq.(\ref{balanced}) eases  the holographic description. The field theories enjoy $SO(1,q-1)$ Lorentz symmetry, and {\it at least} an $SU(2)_R$ global R-symmetry (related to the eight preserved Poincare supercharges). Aside from these symmetries, there are the global flavour and local gauge groups
\begin{equation*}
 G_{\text{global}}=SU(F_1)\times ....\times SU(F_{P}), ~~G_{\text{local}}=SU(N_1)\times ....\times SU(N_P).   
\end{equation*}
When $q>4$, the field theories flow to a SCFT in the UV (this conformal point is what the holographic duals studied below describe). When $q<4$, the SCFTs appear at low energies and it is in the IR that the holographic set-ups described below become valid. For $q=4$ the balancing condition (together with the presence of eight supercharges) guarantees conformality. In all cases, the holographic solutions are trustable in the scaling $P\to\infty$ and $N_i\to\infty$.

In the balanced case, it is useful to describe the quiver in terms of a rank function. This is a convex polygonal function given by
\begin{equation}
     {\cal R}(\eta) = \begin{cases} 
          N_1 \eta & 0\leq \eta \leq 1 \\
          N_l+ (N_{l+1} - N_l)(\eta-l) & l \leq \eta\leq l+1,\;\;\; l:=1,...., P-2\\
  %
          N_{P-1}(P-\eta) & (P-1)\leq \eta\leq P . 
       \end{cases} \label{eq:Rank}
\end{equation}
Note that the rank of the colour groups appear at each integer-value of $\eta=1,2,3,4....$ 
\\Otherwise, the ranks of the flavour groups $SU(F_i)$, appear taking two derivatives of the rank function,
\begin{equation}
{\cal R}''(\eta)= \sum_{i=1}^P F_i \delta(\eta -i).
\end{equation}
The variable $\eta$ is bounded in the interval $[0,P]$ and is associated with the quiver-direction.
It is useful to calculate the (odd) Fourier transform of the rank function in eq.(\ref{eq:Rank}), see \cite{Akhond:2021ffz} for a derivation,
\begin{equation}
 R_k=\frac{2}{P}\int_0^P {\cal R}(\eta)\sin\left( \frac{k\pi \eta}{P}\right) d\eta=\frac{2 P}{\pi^2 k^2}   \sum_{j=1}^{{P-1}}F_j \sin\left(\frac{k\pi j}{P}\right).\label{rankfFourier}
\end{equation}
The holographic description of the field theory (at the strongly coupled conformal point) is written in terms of a function $V$ (sometimes referred to as potential). This function solves a Laplace-like PDE with determined boundary and initial conditions. The  solution to this PDE is written in terms of the Fourier transform of the rank function. All the warp factors in the metric are written in terms of the function $V$ and its derivatives. In other words, the knowledge of the function $V$ is equivalent to knowing the  holographic  description of the CFT. We give some details for each dimension below, and direct the readers to the original references. The logic that we make more explicit in the 3d-case follows similarly in all other dimensions.

The material  is organised in subsections, one for each spacetime dimension. We climb-up from dimension $d=3$ up to $d=6$ SCFTs. The case of dimension $d=2$ is a bit of an outlier and is reserved for the end of the section. In each subsection we write the (relevant part of)  supergravity background, compute the time-like entanglement entropy on slabs and on spherical regions. We also relate the Liu-Mezei and slab central charges with the holographic central charge.
\subsection{Three-dimensional $ N  = 4$ SUSY linear quivers and their holographic dual}
The system has been studied in various papers. Originally in \cite{DHoker:2007hhe, Assel:2011xz}, and in the way described here in the papers \cite{Akhond:2021ffz, Akhond:2022oaf}. The string-frame metric is  
\begin{eqnarray}
   & & ds^2_{10}=f_1 \Bigg[ds^2_{AdS_4}+f_2 d\Omega_2 (\theta, \phi)+ f_3 d \tilde{\Omega}_2 (\tilde{\theta} , \tilde{\phi})+f_4 (d \sigma^2 + d \eta^2)\Bigg],\nonumber\\
    & & ds^2_{AdS_4} = u^2 (\lambda dt^2 + dx^2 + dy^2)+\frac{du^2}{u^2}, ~~~~e^{-4 \Phi}=f^2_5.\label{AdS4configu}
\end{eqnarray}
The function $f_i(\sigma,\eta)$ (the warp factors) are written in terms of a function $V(\sigma,\eta)$
 as \cite{Akhond:2021ffz}
\begin{align}
 & f_1 =  \frac{\pi}{2}\sqrt{\frac{\sigma^3 \partial^2_{\sigma \eta}V}{\partial_{\sigma}(\sigma \partial_\eta V)}},~~ f_2 = -\frac{\partial_\eta V \partial_\sigma (\sigma \partial_\eta V)}{\sigma \Lambda},~~f_3=\frac{\partial_\sigma (\sigma \partial_\eta V)}{\sigma \partial^2_{\sigma \eta} V}\\
 & f_4 = - \frac{\partial_\sigma (\sigma \partial_\eta V)}{\sigma^2 \partial_\eta V},~~ f_5 = -16 \Lambda \frac{\partial_\eta V}{\partial^2_{\sigma \eta } V}, ~~ \Lambda = \partial_\eta V \partial^2_{\sigma \eta}V + \sigma (\partial^2_{\sigma \eta}V)^2 +\sigma (\partial^2_\eta V)^2.\nonumber
\end{align}
Note that we have set the parameter $l=1$ (unit AdS-radius) to avoid cluttering of the notation.
Importantly, the function $V(\sigma,\eta)$ can be written in terms of the rank function (specifically, its Fourier transform) \cite{Akhond:2021ffz},
\begin{equation}
V(\sigma,\eta)=\frac{1}{\sigma}\sum_{k=1}^\infty R_k \cos\left(\frac{k\pi\eta}{P} \right) e^{-\frac{k \pi |\sigma|}{P}}.\label{V3d}
\end{equation}
This illustrates the logic for the construction of the pair holographic background-SCFT: \begin{itemize}
    \item{
first choose a balanced quiver as  in Figure \ref{fig:quiver}.}
\item{Then encode the quiver in the  rank function in eq.(\ref{eq:Rank}).}
\item{After this, compute its (odd) Fourier transform and
finally, with these quantities, write the function $V(\sigma,\eta)$ in eq.(\ref{V3d}) to be replaced in the holographic background in eq.(\ref{AdS4configu}).}
\end{itemize}
Similar expressions exist for the other NS and RR fields. Exactly the same logic applies to other dimensions.
\\
To compute the time-like entanglement entropy,  we work with an eight-manifold parametrised by the coordinates $[x, u,\Omega_2,\tilde{\Omega}_2,\sigma,\eta ]$ with  $t(u)$ and $y=0$. The induced metric is (as above $\lambda=\pm 1$),
\begin{equation}
    ds_8^2 
    = \frac{f_1}{u^2}(1-\lambda u^4 t'^2(u))du^2 +f_1 u^2 dx_1^2+f_1 f_2 d \Omega_2 (\theta , \phi) + f_1 f_3 d \tilde{\Omega}_2 (\tilde{\theta}, \tilde{\phi}) + f_1 f_4 (d \sigma^2 + d \eta^2). 
\end{equation}

The EE that follows from eqs.(\ref{SEEAdSfinal})-(\ref{SEEAdSfinal2}) is
\begin{equation}
 \frac{4G_{10}}{\mathcal{N}} S_{EE}[\Sigma_8] =
    \begin{cases}
      -\frac{4 \pi  \Gamma \left(\frac{3}{4}\right)^2}{\Gamma \left(\frac{1}{4}\right)^2} \frac{1}{|T|}& \text{In Euclidean ($\lambda=+1$) signature}\\
      \frac{4 \pi i \Gamma \left(\frac{3}{4}\right)^2}{\Gamma \left(\frac{1}{4}\right)^2} \frac{1}{|T|} & \text{In Lorentzian ($\lambda=-1$) signature}.
    \end{cases}       
\end{equation}
We have defined,
\begin{equation}
 \mathcal{N}= -16 \pi^6 L_{x_1}\int_0^\infty d\sigma \int_0^P d \eta \sigma^2 \partial_\eta V \partial_\sigma (\sigma \partial_\eta V).   
\end{equation}
The quantity ${\cal N}$ is related to the holographic central charge defined in \cite{Akhond:2021ffz, Uhlemann:2019ypp}. The approximate expressions \eqref{e20}-\eqref{e22}, yield
\begin{equation}
 T_{app} =
    \begin{cases}
      \frac{ \pi }{2u_0} & \text{In Euclidean ($\lambda=+1$) signature}\\
      \frac{-i \pi }{2u_0} & \text{In Lorentzian ($\lambda=-1$) signature}
    \end{cases}       
\end{equation}
\begin{equation}
 S_{EE,approx} =
    \begin{cases}
     - \frac{ \pi^2 }{4} \frac{1}{|T_{app}|}& \text{In Euclidean ($\lambda=+1$) signature}\\
      \frac{ i\pi^2 }{4} \frac{1}{|T_{app}|} & \text{In Lorentzian ($\lambda=-1$) signature}.
    \end{cases}       
\end{equation}

For the spherical ($\lambda=+1$) or hyperboloid ($\lambda=-1$) regions, from eq.\eqref{e31} we find, 
\begin{equation}
 \frac{4G_{10}}{\hat{\mathcal{N}}}S^{(\lambda)}_{EE}[{\Sigma}_8]=
    \begin{cases}
     \frac{R}{\epsilon}-1& \text{In Euclidean ($\lambda=+1$) signature}\\
      \frac{R}{\epsilon}-i & \text{In Lorentzian ($\lambda=-1$) signature}.
    \end{cases}       
\end{equation}
We have defined
\begin{equation}
  \hat{\mathcal{N}}=  -16 \pi^6 \text{Vol}(\Omega^{(\lambda)}_{1})\int_0^\infty d\sigma \int_0^P d \eta\sigma^2 \partial_\eta V \partial_\sigma (\sigma \partial_\eta V).
\end{equation}
We compute the Liu-Mezei and slab central charges \cite{Liu:2012eea} using the definitions in eqs. \eqref{e33} and \eqref{e36}, 
\begin{align}
    c_{LM}=\frac{\hat{\mathcal{N}}}{4G_{10}}~,~c_{slab}=\frac{\mathcal{N}}{4G_{10}}\frac{\kappa}{L}\frac{4 \pi  \Gamma \left(\frac{3}{4}\right)^2}{\Gamma \left(\frac{1}{4}\right)^2},
\end{align}
where only the \emph{absolute} values are considered. It is interesting to observe that
both these quantities and the holographic central charge \cite{Akhond:2021ffz} are proportional
\begin{equation}
 c_{LM}\propto c_{slab}\propto c_{hol}\sim \sum_{k=1}^\infty k R_k^2.   
\end{equation}
These three quantities and the EE are U-duality  (T-duality, S-duality and non-abelian T-duality \cite{Macpherson:2014eza}) and also mirror symmetry invariants. The Liu-Mezei, slab and holographic central charges have different normalisations (they are only proportional to each other). What is interesting is that they are sensitive to the same quantity, represented here by $\sum_{k=1}^\infty k R_k^2$. This is a characteristic quantity for each linear quiver SCFT$_3$. Similar comments apply to other dimensions.
\\
Let us now study the case of an infinite family of four dimensional SCFTs.
\subsection{Four-dimensional $N  = 2$ SUSY linear quivers and their holographic dual}
We study $AdS_5$ backgrounds in Type IIA, dual to $N=2$ SCFTs of the linear quiver type. These backgrounds were discussed by Gaiotto and Maldacena in \cite{Gaiotto:2009gz}, see also \cite{Reid-Edwards:2010vpm, Aharony:2012tz, Lozano:2016kum, Macpherson:2024frt, Nunez:2018qcj, Nunez:2019gbg} for further elaborations. The corresponding string-frame metric and dilaton are,
\begin{align}
    &ds^2_{10}=\sqrt{f^3_1 f_5} \Bigg[ 4ds^2_{AdS_5}+f_2 d\Omega_2 (\theta , \phi)+ f_3 d \chi^2 +f_4 (d \sigma^2 + d \eta^2)\Bigg]\\
    & ds^2_{AdS_5} = u^2 (\lambda dt^2 + dx^2_1 + dx^2_2+dy^2)+\frac{du^2}{u^2}\\
    &e^{-4 \Phi}=(f_1 f_5)^{-3}.
\end{align}
Again, we allow $\lambda=\pm 1$ and we set the AdS-scale $l=1$.
The functions $f_i(\sigma,\eta)$ are written in terms of derivatives of the potential function $V(\sigma,\eta)$. The expressions are \cite{Nunez:2018qcj, Macpherson:2024frt}
\begin{align}
    &f^3_1 = \frac{\dot{V}\Delta}{2V''},f_2=\frac{2 V'' \dot{V}}{\Delta},f_3=\frac{4 \sigma^2 V''}{2\dot{V}-\ddot{V}},f_4=\frac{2 V''}{\dot{V}},f_5=\frac{2(2 \dot{V}-\ddot{V})}{\dot{V} \Delta}\\
    & \Delta = (2 \dot{V}-\ddot{V})V''+(\dot{V}')^2, \dot{V}=\sigma \partial_\sigma V, V'' =\partial^2_\eta V
\end{align}
where we denote $\dot{V}=\sigma \partial_\sigma V$ and $V'=\partial_\eta V$.
The potential function solves a linear PDE \cite{Nunez:2019gbg, Macpherson:2024frt} whose solution is
\begin{equation}
V(\sigma,\eta)=-\sum_{k=1}^\infty R_k \sin\left( \frac{k\pi\eta}{P}\right) K_0\left( \frac{k\pi \sigma}{P}\right).   
\end{equation}
The reasoning from the three-dimensional case applies here as well.
To compute the EE, we choose the eight-manifold to be $\Sigma_8=[x_1,x_2, u, \Omega_2,\chi,\sigma,\eta]$, setting $y=0$ and considering an embedding $t=t(u)$. The time-like entanglement entropy for the slab follows from eqs. \eqref{SEEAdSfinal}-\eqref{SEEAdSfinal2} setting $Y=0$, 
\begin{equation}
 \frac{4G_{10}}{\mathcal{N}} S_{EE}[\Sigma_8^{(\lambda)}] =
    \begin{cases}
      -\frac{4 \pi ^{3/2} \Gamma \left(\frac{2}{3}\right)^3}{\Gamma \left(\frac{1}{6}\right)^3}\frac{1}{|T|^2}& \text{In Euclidean ($\lambda=+1$) signature}\\
      \frac{4 \pi ^{3/2} \Gamma \left(\frac{2}{3}\right)^3}{\Gamma \left(\frac{1}{6}\right)^3}\frac{1}{|T|^2} & \text{In Lorentzian ($\lambda=-1$) signature}
    \end{cases}       
\end{equation}
which differ only by an overall sign. Also, we have defined the constant,
\begin{equation}
 \mathcal{N}= 256 \pi^2 L_{x_1}L_{x_2}\int_0^\infty d \sigma \int_0^P d\eta ~\sigma \dot{V}V''. \label{calNparaGM}  
\end{equation}
It was found in \cite{Nunez:2019gbg, Macpherson:2024frt}, that the holographic central charge is proportional to ${\cal N}$.

The approximate expressions for the time separation and the tEE in eqs. \eqref{e20}-\eqref{e22} are
\begin{equation}
 T_{app} =
    \begin{cases}
      \frac{ \pi }{3u_0} & \text{In Euclidean ($\lambda=+1$) signature}\\
      \frac{-i \pi }{3u_0} & \text{In Lorentzian ($\lambda=-1$) signature}
    \end{cases}       
\end{equation}
\begin{equation}
 S_{EE,approx} =
    \begin{cases}
     -\frac{\pi ^3}{54} \frac{1}{|T_{app}|^2}& \text{In Euclidean ($\lambda=+1$) signature}\\
      \frac{\pi ^3}{54} \frac{1}{|T_{app}|^2} & \text{In Lorentzian ($\lambda=-1$) signature}.
    \end{cases}       
\end{equation}
The EE for the sphere/hyperboloid region can be computed from eq.\eqref{e32}, 
\begin{equation}
 \frac{4G_{10}}{\hat{\mathcal{N}}}S_{EE}[\hat{\Sigma}_8^{(\lambda)}]=
    \begin{cases}
     \frac{R^2}{2 \epsilon^2}-\frac{1}{2}\log (\frac{2R}{\epsilon })-\frac{1}{4}& \text{In Euclidean ($\lambda=+1$) signature}\\
     \frac{R^2}{2 \epsilon^2}+\frac{1}{2}\log (\frac{2R}{\epsilon })+\frac{1}{4}(1-i\pi) & \text{In Lorentzian ($\lambda=-1$) signature}.
    \end{cases}       
\end{equation}
We have defined the constant 
\begin{equation}
 \hat{\mathcal{N}}=256 \pi^2 \text{Vol}(\Omega^{(\lambda)}_{2})\int_0^\infty d \sigma \int_0^P d\eta ~\sigma \dot{V}V''.   
\end{equation}
With these results, the Liu-Mezei and slab central charges \cite{Liu:2012eea},\cite{Jokela:2025cyz} follow using eqs.\eqref{e34} and \eqref{e36}, 
\begin{align}
    c_{LM}=\frac{\hat{\mathcal{N}}}{8G_{10}}~,~c_{slab}=\frac{\mathcal{N}}{4G_{10}}\frac{\kappa}{L_{x_1}L_{x_2}}\frac{8 \pi ^{3/2} \Gamma \left(\frac{2}{3}\right)^3}{\Gamma \left(\frac{1}{6}\right)^3}.
\end{align}
We have considered only the absolute value. 
In this case we find that
\begin{equation}
 c_{LM}\propto c_{slab}\propto c_{hol}\sim P \sum_{k=1}^\infty  R_k^2.   
\end{equation}
The Fourier transform $R_k$ is calculated as in eq.(\ref{rankfFourier}).
Let us now study the situation for an infinte family of AdS$_6$ solutions in Type IIB supergravity.
 \subsection{Five-dimensional $ N  = 1$ SUSY linear quivers and their holographic dual}
We consider an infinite family of Type IIB solutions containing an $AdS_6$ factor \cite{DHoker:2016ysh, DHoker:2016ujz, Apruzzi:2018cvq, Legramandi:2021uds, Legramandi:2021aqv}. These backgrounds are dual to $ N  = 1$ linear quivers in five-dimensions. In the formalism we use here, the supergravity backgrounds are \cite{Legramandi:2021uds}
\begin{align}
    &ds^2_{10}=f_1 \Bigg[ds^2_{AdS_6}+f_2 d\Omega_2 (\theta , \phi) + f_3 (d \sigma^2 + d \eta^2)\bigg]\\
    & ds^2_{AdS_6} = u^2 (\lambda dt^2 + dx^2_1 + dx^2_2+dx^2_3+dy^2)+\frac{du^2}{u^2}\\
    &e^{-4 \Phi}=f^2_6
\end{align}
As above, we have set the AdS-scale $l=1$ and kept the parameter $\lambda=\pm 1$. The functions $f_i(\sigma,\eta)$ are written in terms of a potential $V(\sigma,\eta)$ and its derivatives, 
\begin{align}
    &f_1 = \frac{2}{3}\Big(\frac{\sigma(\sigma \partial^2_\eta V+ 3 \partial_\sigma V)}{\partial^2_\eta V} \Big)^{1/2},~~f_2= \frac{\partial_\sigma V \partial^2_\eta V}{3 \tilde{\Lambda}}, ~~f_3=\frac{\partial^2_\eta V}{3 \sigma \partial_\sigma V}\\
    &f_6 = 18^2 \frac{3 \sigma^2 \partial_\sigma V \partial^2_\eta V \tilde{\Lambda}}{(3 \partial_\sigma V + \sigma \partial^2_\eta V)^2},~~\tilde{\Lambda}=\sigma (\partial^2_{\sigma \eta}V)^2+\partial^2_\eta V (\partial_\sigma V - \sigma \partial^2_\sigma V).
\end{align}
The function $V(\sigma,\eta)$ satisfies a linear PDE \cite{Legramandi:2021aqv}, whose solution is
\begin{equation}
 V(\sigma,\eta)=-\frac{1}{\sigma}\sum_{k=1}^\infty \frac{P}{2\pi k} R_k \sin \left( \frac{k\pi \eta}{P}\right) e^{-\frac{k \pi |\sigma|}{P}} . 
\end{equation}
The calculation of the timelike entanglement entropy requires us to define an eight-manifold $\Sigma_8=[x_1,x_2,x_3, u, \Omega_2,\sigma,\eta]$ with $y=0$ and $t(u)$. We find,
\begin{equation}
 \frac{4G_{10}}{\mathcal{N}} S_{EE}[\Sigma_8^{(\lambda)}] =
    \begin{cases}
      -\frac{16 \pi ^2 \Gamma \left(\frac{5}{8}\right)^4}{3 \Gamma \left(\frac{1}{8}\right)^4}\frac{1}{|T|^3}& \text{In Euclidean ($\lambda=+1$) signature}\\
      -\frac{16 i \pi ^2 \Gamma \left(\frac{5}{8}\right)^4}{3 \Gamma \left(\frac{1}{8}\right)^4}\frac{1}{|T|^3} & \text{In Lorentzian ($\lambda=-1$) signature}.
    \end{cases}       
\end{equation}
We have defined
\begin{equation}
\mathcal{N}= \frac{2^8 \pi}{3} L_{x_1}L_{x_2}L_{x_3}\int_0^\infty d\sigma \int_0^P d \eta~~ \sigma^3 \partial_\sigma V \partial^2_\eta V %
\end{equation}
The approximate expressions in eqs. \eqref{e20}-\eqref{e22} read, 
\begin{equation}
 T_{app} =
    \begin{cases}
      \frac{ \pi }{4u_0} & \text{In Euclidean ($\lambda=+1$) signature}\\
      \frac{-i \pi }{4u_0} & \text{In Lorentzian ($\lambda=-1$) signature}
    \end{cases}       
\end{equation}
\begin{equation}
 S_{EE,approx} =
    \begin{cases}
    -\frac{\pi ^4}{768} \frac{1}{|T_{app}|^3}& \text{In Euclidean ($\lambda=+1$) signature}\\
      -\frac{i \pi ^4}{768}  \frac{1}{|T_{app}|^3} & \text{In Lorentzian ($\lambda=-1$) signature}.
    \end{cases}       
\end{equation}
For the sphere/hyperboloid regions, the time-like entanglement entropy can be computed using eq. \eqref{e32}, 
\begin{equation}
 \frac{4G_{10}}{\hat{\mathcal{N}}}S_{EE}[\hat{\Sigma}_8^{(\lambda)}]=
    \begin{cases}
     \frac{R^3}{3 \epsilon^3}- \frac{R}{\epsilon}+\frac{2}{3}& \text{In Euclidean ($\lambda=+1$) signature}\\
    \frac{R^3}{3 \epsilon^3}+ \frac{R}{\epsilon}-i\frac{2}{3} & \text{In Lorentzian ($\lambda=-1$) signature}.
    \end{cases}       
\end{equation}
We defined,
\begin{equation}
\hat{\mathcal{N}}=  \frac{2^8 \pi}{3} \text{Vol}(\Omega^{(\lambda)}_{3})\int_0^\infty d\sigma \int_0^P d \eta~~\sigma^3 \partial_\sigma V \partial^2_\eta V.    
\end{equation}
The Liu-Mezei central charge in eq.\eqref{e33} and the slab central charge in eq. \eqref{e36} are,
\begin{align}
    c_{LM}=\frac{\hat{\mathcal{N}}}{4G_{10}}\frac{2}{3}~,~c_{slab}=\frac{\mathcal{N}}{4G_{10}}\frac{\kappa}{L_{x_1}L_{x_2}L_{x_3}}\frac{16 \pi ^2 \Gamma \left(\frac{5}{8}\right)^4}{\Gamma \left(\frac{1}{8}\right)^4}.
\end{align}
As it happens in other cases and using $R_k$ defined in eq.(\ref{rankfFourier}), we find
\begin{equation}
 c_{LM}\propto c_{slab}\propto c_{hol}\sim P^2 \sum_{k=1}^\infty \frac{ R_k^2}{k}.   
\end{equation}
\subsection{Six-dimensional $ N  = (1,0)$ SUSY linear  quivers and their holographic dual}
We consider massive Type IIA supergravity backgrounds with an $AdS_7$ factor \cite{Apruzzi:2015wna, Apruzzi:2014qva,Cremonesi:2015bld, Nunez:2018ags, Filippas:2019puw}. These are conjectured to be dual to $N$=(1,0) SCFTs in six-dimensions. The corresponding string-frame metric reads \cite{Nunez:2018ags},\cite{Filippas:2019puw}
\begin{align}
    &ds^2_{10}=f_1 ds^2_{AdS_7}+f_2 d\eta^2 +f_3 d\Omega_2 (\theta , \phi) \\
    & ds^2_{AdS_6} = u^2 (\lambda dt^2 + dx^2_1 + dx^2_2+dx^2_3+dx^2_4 + dy^2)+\frac{du^2}{u^2}\\
    &e^{-4 \Phi}=f^{-4}_6.
\end{align}
As before, we set the AdS-scale $l=1$ and keep the parameter $\lambda=\pm1$. The functions $f_i(\eta)$ can be written in terms of a potential $V(\eta)$ as 
\begin{align}
    &f_1 = 8 \sqrt{2}\pi \sqrt{-\frac{V}{V''}},~~~~~~ f_2= \sqrt{2}\pi \sqrt{- \frac{V''}{V}},\\
    &f_3=f_2 \frac{V^2}{(V'^2 - 2 V V'')},~~~f_6^4 = \frac{2^5 \pi^{10}3^{16}\Big(-\frac{V}{V''} \Big)^3}{(V'^2 - 2 V V'')^2}.
\end{align}
The potential function $V(\eta)$ satisfies a linear ODE \cite{Cremonesi:2015bld,Nunez:2018ags, Filippas:2019puw},
\begin{equation}
 V'''=-162 \pi^3 F_0.   
\end{equation}
Being $F_0$ a piece-wise constant RR zero form. By (even) Fourier expanding $F_0$, we find a Fourier expansion of $V'''$. By integration, we have the (odd)-Fourier expansion of $V(\eta)$. In this case, the role of the rank function is played by $V''(\eta)$. 

To compute the time like entanglement entropy, we set an eight manifold $\Sigma_8=[x_1,x_2,x_3,x_4, \Omega_2, u,\eta]$ with $y=0$ and $t(u)$.
The time-like EE follows from eqs.(\ref{SEEAdSfinal})-(\ref{SEEAdSfinal2}),
\begin{equation}
 \frac{4G_{10}}{\mathcal{N}} S_{EE}[\Sigma_8^{(\lambda)}] =
    \begin{cases}
      -\frac{8 \pi ^{5/2} \Gamma \left(\frac{3}{5}\right)^5}{\Gamma \left(\frac{1}{10}\right)^5}\frac{1}{|T|^4}& \text{In Euclidean ($\lambda=+1$) signature}\\
      -\frac{8 \pi ^{5/2} \Gamma \left(\frac{3}{5}\right)^5}{\Gamma \left(\frac{1}{10}\right)^5}\frac{1}{|T|^4} & \text{In Lorentzian ($\lambda=-1$) signature}
    \end{cases}       
\end{equation}
which is interestingly the same in both signatures.
We have defined
\begin{equation}
\mathcal{N}= \frac{4}{3}\Big( \frac{2}{3}\Big)^7 L_{x_1}....L_{x_4}\int_0^P (- V'' V) d\eta
\end{equation}
The approximate expressions in eqs.\eqref{e20}-\eqref{e22} read,
\begin{equation}
 T_{app} =
    \begin{cases}
      \frac{ \pi }{5u_0} & \text{In Euclidean ($\lambda=+1$) signature}\\
      \frac{-i \pi }{5u_0} & \text{In Lorentzian ($\lambda=-1$) signature}
    \end{cases}       
\end{equation}
\begin{equation}
 S_{EE,approx} =
    \begin{cases}
     -\frac{\pi ^5}{12500}\frac{1}{|T_{app}|^4}& \text{In Euclidean ($\lambda=+1$) signature}\\
      -\frac{\pi ^5}{12500}\frac{1}{|T_{app}|^4} & \text{In Lorentzian ($\lambda=-1$) signature}.
    \end{cases}       
\end{equation}

The time-like entanglement for the sphere/hyperboloid regions is computed using eq. \eqref{e32}, 
\begin{equation}
 \frac{4G_{10}}{\hat{\mathcal{N}}}S^{(\lambda)}_{EE}[\hat{\Sigma}_8^{(\lambda)}]=
    \begin{cases}
     \frac{R^4}{4 \epsilon^4}-\frac{3}{4}\frac{R^2}{\epsilon^2}+ \frac{3}{8} \left(\log (\frac{2R}{\epsilon})+\frac{3}{4}\right)& \text{In Euclidean ($\lambda=+1$) signature}\\
     \frac{R^4}{4 \epsilon^4}+\frac{3}{4}\frac{R^2}{\epsilon^2}+ \frac{3}{8} \left(\log (\frac{2R}{\epsilon })-\frac{i \pi}{2}+\frac{3}{4}\right) & \text{In Lorentzian ($\lambda=-1$) signature}.
    \end{cases}       
\end{equation}
We have defined
\begin{equation}
 \hat{\mathcal{N}}=  \frac{4}{3}\Big( \frac{2}{3}\Big)^7 \text{Vol}(\Omega^{(\lambda)}_{4})\int_0^P (- V'' V) d\eta.   
\end{equation}
We compute the Liu-Mezei central charge \cite{Liu:2012eea} and the slab central charge \cite{Jokela:2025cyz}. These follow from eqs.\eqref{e34} and \eqref{e36}. The results are,
\begin{align}
    c_{LM}=\frac{\hat{\mathcal{N}}}{4G_{10}}\frac{3}{8}~,~c_{slab}=\frac{\mathcal{N}}{4G_{10}}\frac{\kappa}{L_{x_1}....L_{x_4}}\frac{32 \pi ^{5/2} \Gamma \left(\frac{3}{5}\right)^5}{\Gamma \left(\frac{1}{10}\right)^5}.
\end{align}
As found before, using $R_k$ defined in eq.(\ref{rankfFourier}), in this case we have
\begin{equation}
 c_{LM}\propto c_{slab}\propto c_{hol}\sim P^3 \sum_{k=1}^\infty  \frac{R_k^2}{k^2}.   
\end{equation}
%



\subsection{Two-dimensional $N=(0,4)$ SUSY quivers and holographic dual}
We briefly study here the case of AdS$_3$ here.
%
%
%
%
We present this case last, as it differs somewhat from the material above. The differences stem from two facts: first that the dual field theories are not simple linear quivers, but 'two lines' quivers with many bifundamental fields connecting various nodes (hence, two rank functions are used). For details of the field theories, see \cite{Lozano:2019zvg, Lozano:2019ywa, Couzens:2021veb}. The supergravity configuration is written in \cite{Lozano:2019zvg, Lozano:2019jza,Lozano:2019emq}. The second difference is that the generic treatment in eqs(\ref{Tads})-(\ref{SEEAdSfinal2}), should be carefully handled. In fact, powers of $(d-2)$ should translate into logarithms as we discuss below. Let us go over these in some detail.

The metric and  dilaton (we omit the Ramond and Neveu-Schwarz fields), are
\begin{align}
ds^2&= \frac{\hat{G}}{\sqrt{\hat{h}_4 h_8}}\bigg(ds^2(\text{AdS}_3)+\frac{h_8\hat{h}_4 }{4 h_8\hat{h}_4+(\hat{G}')^2}ds^2(\text{S}^2)\bigg)+ \sqrt{\frac{\hat{h}_4}{h_8}}ds^2(\text{CY}_2)+ \frac{\sqrt{\hat{h}_4 h_8}}{\hat{G}} d\rho^2,\\
e^{-\Phi}&= \frac{h_8^{\frac{3}{4}} }{2\hat{h}_4^{\frac{1}{4}}\sqrt{\hat{G}}}\sqrt{4h_8 \hat{h}_4+(\hat{G}')^2}.~~~ds^2(\text{AdS}_3)=\frac{u^2}{l^2}(-dt^2+dy^2)+\frac{l^2 du^2}{u^2}\nonumber
\end{align}
The functions $\hat{h}_4, h_8$ and $\hat{G}$ depend only on the coordinate $\rho$ (that ranges in the $[0,P]$-interval) and solve second order linear ODEs. For the details see \cite{Lozano:2019zvg, Lozano:2019jza}.

The eight-manifold needed to calculate the entanglement entropy is defined by $\Sigma_8=[u, \Omega_2,CY_2,\rho]$ with $t(u), y(u)$. The induced metric is (we take $\lambda=-1$, the Lorentzian signature),
\begin{eqnarray}
& &   ds^2_{\Sigma_8}= \frac{\hat{G}}{\sqrt{\hat{h}_4 h_8}}\bigg(du^2\left[\frac{l^2}{u^2}+\frac{u^2}{l^2}(y'^2-t'^2)\right]+\frac{h_8\hat{h}_4 }{4 h_8\hat{h}_4+(\hat{G}')^2}ds^2(\text{S}^2)\bigg)+ \sqrt{\frac{\hat{h}_4}{h_8}}ds^2(\text{CY}_2)+ \frac{\sqrt{\hat{h}_4 h_8}}{\hat{G}} d\rho^2,\nonumber\\
& & e^{-4\Phi}\det[g_{\Sigma_8}]=\frac{\hat{h}_4^2 h_8^2}{16}\text{Vol}_{S^2}\text{Vol}_{CY_2}\left[\frac{l^2}{u^2}+\frac{u^2}{l^2}(y'^2-t'^2)\ \right].
\end{eqnarray}
The entanglement entropy is
\begin{eqnarray}
& &\frac{4G_{10}}{\cal N} S_{EE}= \int_{u_0}^\infty du \sqrt{\frac{l^2}{u^2}+\frac{u^2}{l^2}(y'^2-t'^2)},\label{EEAdS3}\\
& &
{\cal N}=
\pi \text{Vol}_{CY_2}\int_0^P \hat{h}_4 h_8d\rho .\label{EEAdS32}  
\end{eqnarray}
The expressions for the separations $T,Y$ and the regulated EE are,
\begin{eqnarray}
& &\frac{T}{2c_t l^3}=\frac{Y}{2c_y l^3}=\int_{u_0}^\infty \frac{du}{u^2\sqrt{u^2-u_0^2}}= \frac{1}{u_0^2},\label{Tads3exact}\\
& &\frac{4G_{10}}{\cal N}S_{EE}=\lim_{\Lambda\to\infty}\int_{u_0}^\Lambda\frac{du}{\sqrt{u^2-u_0^2}} -\int_{\frac{1}{\Lambda}}^\Lambda \frac{du}{u}=\log\left[\frac{\sqrt{Y^2-T^2}}{\Lambda l^3} \right].\label{EESEPAdS3}
\end{eqnarray}
Using the functions $G(u)=\frac{l}{u}$ and $F(u)=\frac{u}{l}$, and the approximate expressions in eqs.(\ref{e20})-(\ref{e21}), we find the same functional dependence of $T_{app}$ and $S_{app}$ in terms of $u_0$ as that in eqs.(\ref{Tads3exact})-(\ref{EESEPAdS3}). 

Notice that in this case, the holographic central charge (computed in field theory and matching the gravity calculation, see \cite{Lozano:2019jza}) is proportional to ${\cal N}$ defined in eq.(\ref{EEAdS32}) which coincides with the Liu-Mezei central charge.

Let us now study qualitatively different physical systems. Below, we discuss systems with a scale, which makes the field theory gapped and in some cases,  confining. Our aim is to calculate the timelike entanglement entropy in these cases, extending the results of \cite{Chu:2019uoh, Afrasiar:2024lsi, Nunez:2025gxq, Nunez:2025ppd}.

\section{Study of gapped and/or confining models}\label{sectionconfiningmodels}
%
%
%
%
In this section, we extend the above analysis to systems that exhibit a mass gap (and in some cases confinement). We work with class II and class III backgrounds, as defined in eqs.(\ref{classii})-(\ref{classiii}). 

As discussed in \cite{Anabalon:2021tua,Anabalon:2022aig, Anabalon:2024che, Anabalon:2024qhf, Nunez:2023xgl, Fatemiabhari:2024aua,Chatzis:2024kdu,Chatzis:2024top, Chatzis:2025dnu, Giliberti:2024eii, Macpherson:2025pqi} it is possible to construct backgrounds dual to confining field theories, starting with a supersymmetric AdS-background and performing a twisted compactification on a circle, aided by a one-form that mixes the space-circle with a $U(1)_R$ inside the R-symmetry. This type of construction was studied from the QFT viewpoint in \cite{Komargodski:2011vj,Kumar:2024pcz} and  geometrically in \cite{Macpherson:2024qfi}. These backgrounds feature a circle that is fibered over the internal manifold and shrinks smoothly, leading to a gapped dual, in the style of \cite{Witten:1998zw}, but preserving SUSY. We first discuss these systems, summarised in a  metric and dilaton of class II, eq.(\ref{classii}). After this, we move to systems in class III, eq.(\ref{classiii}). These systems are complemented by other Ramond and Neveu-Schwarz fields that we do not quote as they are not needed in the computations of this work.

As we indicated in eq.(\ref{class23-bis}), we define an eight manifold, calculate
the U-duality invariant $e^{-4\Phi}\det[g_{\Sigma_8}]$ and then the entanglement entropy on a slab, as in eq.(\ref{SEEwork}). Below we present different models in the bibliography and write the EE for each of them.

{
Before we proceed further, it is important to highlight the main difference between the conformal examples \cite{ Heller:2024whi,Doi:2023zaf,Nunez:2025gxq} studied in the previous section and the confining models of the present section. For conformal theories, the extremal surface has a turning point characterised by eq.(\ref{u0gen}). In contrast, for the case of confining theories, the extremal surface is characterized by a turning point of the type  $u_0\sim (\Lambda+ c_y^2-c_t^2)^{\frac{1}{2(d-1)}}$,  where $\Lambda$ is the confinement scale. We discuss a special case of this in eq. \eqref{u0conf} below. Referring to Figure \ref{figLC} and the discussion under the figure, one could have  a real turning point, even for pure time-like separated events ($c_y=0$) if $\Lambda>c_t^{2}$. However, the limit $\Lambda \rightarrow 0$ is delicate, making the turning point ($u_0$) imaginary and leading to a Type II extremal surface which is the result of AdS \cite{ Heller:2024whi, Nunez:2025gxq}, as discussed in the previous section. In summary, the crucial difference is played by the confinement scale ($\Lambda \neq 0$), which allows a \textit{real} (or Type I) extremal surface even for purely time-like separated events. These are usual RT-like surfaces that one encounters in the case of space-like separated events in AdS. For the case $\Lambda+ c_y^2< c_t^2$ we are back to a Type II surface, with complex turning point.}

\subsection{Class II backgrounds}
Here, we collect the results for class II backgrounds. We start with the simplest example: the background presented by Anabal\'on and Ross in \cite{Anabalon:2021tua}, after this, we discuss a model presented by Anabal\'on, Nastase and Oyarzo \cite{Anabalon:2024che}.
Subsequent subsections study 'decorations' on these kind of models. These decorations lead to more elaborated field theory dynamics.
\subsubsection{The Anabal\'on-Ross model}\label{sectionAR}
We briefly discuss this model as it provides a clean and simple example. A  detailed derivation of the results is given in \cite{Nunez:2025ppd}.
The background consists of a ten-dimensional metric, a constant dilaton ($\Phi = 0$), and a Ramond five-form (omitted here). The metric is,
\begin{eqnarray}
ds_{10}^2 &=& \frac{u^2}{l^2}\left[\lambda dt^2+dy^2+dx^2 + f(u) d\phi^2 \right]+\frac{l^2 du^2}{f(u) u^2}+ l^2 d\tilde{\Omega}_5^2, \label{ARmetric} \\
d\tilde{\Omega}_5^2 &=& d\theta^2+ \sin^2\theta d\psi^2 + \sin^2\theta \sin^2\psi\left(d\varphi_1-A_1 \right)^2+
\sin^2\theta \cos^2\psi\left(d\varphi_2-A_1 \right)^2 +\cos^2\theta \left(d\varphi_3-A_1 \right)^2, \nonumber \\
A_1 &=& Q\left(1-\frac{l^2Q^2}{u^2} \right)d\phi, \quad f(u)= 1-\left(\frac{Ql}{u}\right)^6. \nonumber
\end{eqnarray}
Here, $\lambda = \pm 1$, and we focus on the Lorentzian case $\lambda = -1$. The radial coordinate $u$ ranges over $[u_\Lambda, \infty)$, with $u_\Lambda = Ql$ marking the smooth end of space for a specific period $L_\phi = \frac{1}{3Q}$ of the $\phi$-coordinate.

We consider embeddings of the eight-dimensional surface $\Sigma_8 = [x, \phi, \tilde{\Omega}_5, u]$, with $t = t(u)$ and $y = y(u)$. The induced metric on $\Sigma_8$ is
\begin{eqnarray}
ds^2_8 = \left[\frac{u^2}{l^2}(\lambda t'^2 + y'^2) + \frac{l^2}{u^2 f(u)}\right] du^2 + \frac{u^2}{l^2} dx^2 + \frac{u^2}{l^2} f(u) d\phi^2 + l^2 d\tilde{\Omega}_5^2.
\end{eqnarray}

The entanglement entropy is given by
\begin{eqnarray}
& & S^{(\lambda)}_{EE} = \frac{\hat{\mathcal{N}}}{4G_{10}} \int_{u_0}^\infty \sqrt{G^2(u) + F^2(u)(\lambda t'^2(u) + y'^2(u))}, \label{EEconf} \\
& & G(u) = \frac{u}{l}, \quad F(u) = \frac{u^3}{l^3} \sqrt{f(u)}, \quad \hat{\mathcal{N}} = l^5 L_x L_\phi \text{Vol}(\tilde{S}^5), \quad L_\phi = \frac{1}{3Q}. \nonumber
\end{eqnarray}
Focusing on  the case $\lambda = -1$ and using eqs.(\ref{separationsTY})-(\ref{regulatedEE}), we find for the separations $T$, $Y$ 
\begin{eqnarray}
\frac{T}{2 l^5 c_t} = \frac{Y}{2 l^5 c_y} = \int_{u_0}^{\infty} \frac{u\, du}{\sqrt{u^6 - u_\Lambda^6} \sqrt{u^6 - u_0^6}}. \label{YT-confining}
\end{eqnarray}
The regularised entanglement entropy becomes,
\begin{eqnarray}
\frac{2 l G_{10}}{\hat{\mathcal{N}}} S_{EE} = \int_{u_0}^\infty du \frac{u \sqrt{u^6 - u_\Lambda^6}}{\sqrt{u^6 - u_0^6}} - \int_{u_\Lambda}^\infty u\, du. \label{EEregconf}
\end{eqnarray}
Here, $u_0$ is the turning point, obtained from $F(u_0)^2=c_y^2-c_t^2$,
\begin{align}
u_0 = l (Q^6 + c_y^2 - c_t^2)^{1/6}. \label{u0conf}
\end{align}
For $d = 4$ and $u_\Lambda = 0$, the above expressions reduce to those of the conformal case. Interestingly, confinement (the presence of the scale $u_\Lambda$) allows a real turning point even for $c_t > c_y$, provided $Q > (c_t^2 - c_y^2)^{1/6}$.
In other words, {\it the presence of the confinement scale allows for time-like slabs to present real (Type I) embeddings}.

On the other hand, if $c_t > \sqrt{Q^6 + c_y^2}$, $u_0$ becomes imaginary
\[
u_0 = l e^{i\pi/6} |Q^6 + c_y^2 - c_t^2|^{1/6},
\]
indicating a Type II embedding.
Let us now define,
\begin{eqnarray}
& & r = \frac{u_0}{u}, \quad \gamma = \frac{u_\Lambda}{u_0}, \label{definitionsconf} \\
& & J_1 = \int_0^1 dr \frac{r^3}{\sqrt{(1 - r^6)(1 - \gamma^6 r^6)}}
= \frac{\sqrt{\pi} \Gamma(\frac{5}{3})}{4 \Gamma(\frac{7}{6})} {}_2F_1\left[\frac{1}{2}, \frac{2}{3}, \frac{7}{6}, \gamma^6\right], \nonumber \\
& &J_2 = \int_{\epsilon}^1 \frac{dr}{r^3} \sqrt{\frac{1 - \gamma^6 r^6}{1 - r^6}} = {\cal J}_2 + \frac{1}{2\epsilon^2} + O(\epsilon^4), \nonumber \\
& &{\cal J}_2 = -\frac{\sqrt{\pi} \Gamma(\frac{2}{3})}{2 \Gamma(\frac{1}{6})} {}_2F_1\left[-\frac{1}{2}, -\frac{1}{3}, \frac{1}{6}, \gamma^6\right], \quad J_3 = \int_\epsilon^{1/\gamma} \frac{dr}{r^3} = \frac{1}{2\epsilon^2} - \frac{\gamma^2}{2}. \nonumber
\end{eqnarray}
We find using eqs.(\ref{YT-confining})-(\ref{EEregconf}),
\begin{eqnarray}
c_t = \frac{T u_0^4}{2 J_1 l^5}, \quad c_y = \frac{Y u_0^4}{2 J_1 l^5}, \quad \frac{2 l G_{10}}{\hat{\mathcal{N}} u_0^2} S_{EE} = J_2 - J_3. \label{TY-confining}
\end{eqnarray}

\begin{figure}[h]
\centering
\includegraphics[width=0.8\linewidth]{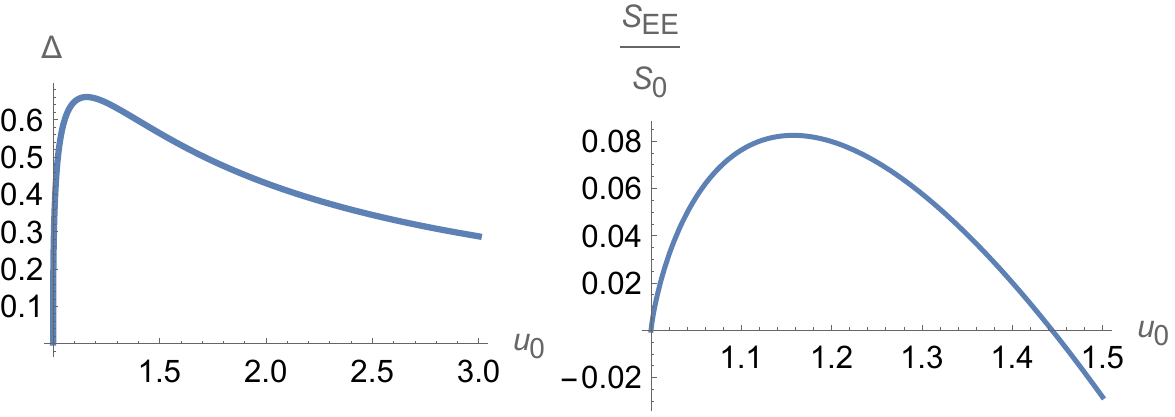}
\caption{Interval $\Delta$ and entanglement entropy vs. $u_0$, with $u_\Lambda = Q = 1$.}
\label{fig1}
\end{figure}

\begin{figure}[h]
\centering
\includegraphics[width=0.5\linewidth]{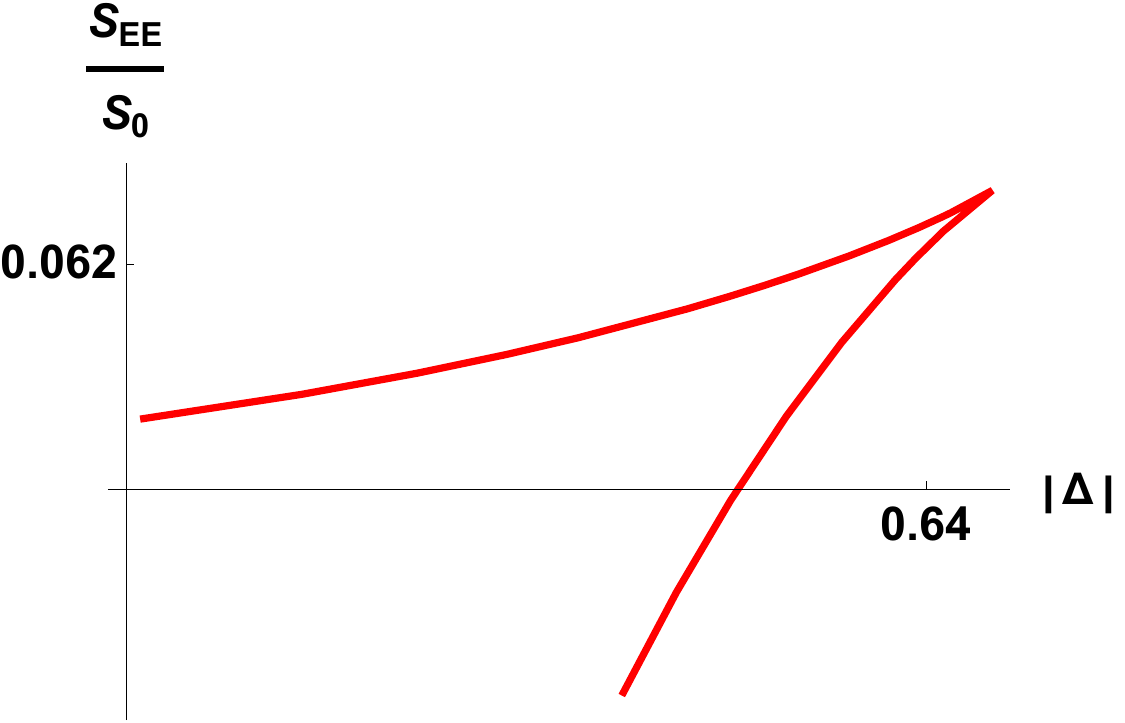}
\caption{Parametric plot of $S_{EE}$ vs. $\Delta$, with $u_\Lambda = Q = 1$. {Here, we set $u_0=1.01$ which corresponds to a real turning point and hence a Type I extremal surface in the bulk. From \eqref{u0conf} one could see that this corresponds to $c_y>c_t$ and hence a \textit{spacelike} separated interval in the dual confining QFTs.}}
\label{fig3}
\end{figure}

Although it is difficult to express $S_{EE}$ explicitly as a function of $\Delta^2 = Y^2 - T^2$, we can write
\begin{eqnarray}
\Delta^2 = (1 - \gamma^6) \frac{4 l^4 J_1^2}{u_0^2}, \quad
S_{EE} = \frac{\hat{\cal N}}{2 l G_N} (J_2 - J_3) u_0^2.\label{SDeltaconf}
\end{eqnarray}
We plot $\Delta^2(u_0)$ and $S_{EE}(u_0)$ in Figure \ref{fig1} and, parametrically, we plot $S_{EE}(\Delta)$ in Figure \ref{fig3}. 

The non-monotonic behavior of $\Delta$ implies a phase transition in $S_{EE}$. Figure \ref{fig3} shows a double-valued $S_{EE}(\Delta)$, a signature of first-order phase transitions in confining theories \cite{Klebanov:2007ws, Kol:2014nqa, Jokela:2020wgs, Jokela:2025cyz, Barbosa:2024pyn}. {Note that this surface is real (Type I surface). Indeed,  $u_0$ is real (we have chosen $Q=1$ and $c_y>c_t$ in Figure \ref{fig3})}. Let us now analyse a related system.
\subsubsection{A Coulomb branch flow. The background of Anabal\'on, Nastase and Oyarzo }\label{coulombbranchgapped}
In this section we study the entanglement entropy for a more elaborate system. The dynamics is that of the Coulomb branch of ${ N}=4$ Super-Yang-Mills \cite{Freedman:1999gk,Freedman:1999gp}. The main difference is that we avoid the characteristically singular behaviour of the supergravity dual, by adding a mass gap to the field theory. In other words, the space ends in a smooth way, after compactification on a shrinking and twisted $S^1$ like it was done in \cite{Anabalon:2024che}. In this case the dilaton is a constant  (we choose $\Phi=0$), and the metric reads\footnote{Note that for this background, the parameter $\lambda=\pm1$ is set to $\lambda=-1$. We also introduce the {\it function} $\lambda(u)$. We hope that this does not cause confusion.}
\begin{equation}\label{S5-1}
    \begin{split}
        \mathrm{d}s^2&=\frac{\zeta(u,\theta)}{L^2}\left[ u^2(-\mathrm{d}t^2+\mathrm{d}y^2+\mathrm{d}x^2+L^2\tilde{f}(u)\mathrm{d}\phi^2)+\frac{L^2\mathrm{d}u^2}{\tilde{f}(u)u^2\lambda^6(u)}+L^4\mathrm{d}\theta^2 \right]\\
        & +\frac{L^2}{\zeta(u,\theta)}\left[ \cos^2\theta\mathrm{d}\psi^2+\cos^2\theta\sin^2\psi \mathrm{D}\phi_1^2 + \cos^2\theta\cos^2\psi \mathrm{D}\phi_2^2 + \lambda^6(u)\sin^2\theta\mathrm{D}\phi_3^2\right],
    \end{split}
\end{equation}

where $\mathrm{D}\phi_i=\mathrm{d}\phi_i+\frac{A^i}{L}$, with
\begin{equation}
 A^1 = A^2 = q_1\left[\lambda^6(u)-\lambda^6(u_{\star})\right]L\, \mathrm{d}\phi,\quad A^3 = q_2\left[\frac{1}{\lambda^6(u)} - \frac{1}{\lambda^6(u_{\star})}\right]L\, \mathrm{d}\phi.\label{one-forms-ANO}  
\end{equation}
The functions $\zeta(u,\theta)$, $\lambda(u)$ and $\tilde{f}(u)$ are defined as,
\begin{eqnarray}
& & \zeta(u , \theta) = \sqrt{1 + \varepsilon \frac{\ell^2}{u^2} \cos^2\theta} ,~~~\lambda^6(u)=\frac{u^2+\varepsilon\ell^2}{u^2},\nonumber\\
& &\tilde{f}(u) =\frac{1}{L^2}-\frac{\varepsilon\ell^2 L^2}{u^4}\left(q_1^2 -\frac{q_2^2}{\lambda^6(u)}\right).\label{functionsANO}
\end{eqnarray}
Here $u_{\star}$ is defined to be the largest root of $\tilde{f}(u)$, satisfying $\tilde{f}(u_{\star})=0$, which expresses the end of the geometry. The quantity $\ell$ is the parameter allowing us to explore the Coulomb branch of the UV-CFT, and $\varepsilon=\pm 1$ is just a sign indicating two non-diffeomorphic branches of the supergravity solution \cite{Anabalon:2024che}.
\\
Together with a Ramond five form, the background is a solution to the supergravity equations of motion. It fits within class II of backgrounds in eq.(\ref{classii}).
To calculate the EE we choose $\Sigma_8=[x,\phi,u,\theta,\psi,\phi_1,\phi_2,\phi_3]$. The quantities needed to write the entanglement entropy are,
\begin{eqnarray}
& & ds_{\Sigma_8}^2=
\frac{\zeta(u,\theta)}{L^2}\Bigg[ u^2( \mathrm{d}x^2+L^2\tilde{f}(u)\mathrm{d}\phi^2)+\frac{L^2\mathrm{d}u^2}{\tilde{f}(u)u^2\lambda^6(u)}\left[1+ \frac{\tilde{f}(u)\lambda^6(u) u^4}{L^2}(-t'^2+ y'^2)\right]+L^4\mathrm{d}\theta^2 \Bigg]\nonumber\\
&        & ~~+\frac{L^2}{\zeta(u,\theta)}\left[ \cos^2\theta\mathrm{d}\psi^2+\cos^2\theta\sin^2\psi \mathrm{D}\phi_1^2 + \cos^2\theta\cos^2\psi \mathrm{D}\phi_2^2 + \lambda^6(u)\sin^2\theta\mathrm{D}\phi_3^2\right],\label{metric8CB}\\
& & e^{-4\Phi}\det[g_{\Sigma_8}]=L^8 \cos^6\theta \sin^2\theta\cos^2\psi\sin^2\psi\left[ u^2+ \frac{\tilde{f}(u) \lambda^6(u) u^6}{L^2}(y'^2-t'^2)\right].\nonumber
\end{eqnarray}
The entanglement entropy reads,
\begin{equation}
\frac{4 G_{10}}{{\cal N}_{II}}S_{EE}=\int_{u_0}^\infty du~ \sqrt{G^2(u)+ F^2(u) (y'^2-t'^2)} , \label{EEforANO}  
\end{equation}
where we have defined,
\begin{eqnarray}
& &    G(u)=u,~~F(u)=\frac{u^3 \lambda^3(u) \sqrt{\tilde{f}(u)}}{L},~~\text{and}\label{FGCoulomb}\\
& & {\cal N}_{II}=L^4 L_x L_\phi \int d\theta d\psi d\phi_1 d\phi_2d\phi_3 \cos^3\theta\sin\theta \cos\psi\sin\psi .\nonumber
\end{eqnarray}
Before writing the explicit expressions for the $T$ and $Y$ separations and the entanglement entropy, let us be more explicit about the functions entering the calculation. We focus on the SUSY case, which implies $q_1=q_2= \frac{q^3 L}{\ell^2}$. In this case we have
\begin{eqnarray}
& & G(u)=u, ~~~\tilde{f}(u)=\frac{\left[ u^6+\epsilon \ell^2 u^4 -L^6 q^6  \right]}{L^2 u^4 (u^2+\epsilon \ell^2)}  ,\nonumber\\
& & F(u)= \frac{1}{L^2}\left[u^6 +\epsilon \ell^2 u^4 - L^6 q^6 \right]^{\frac{1}{2}}.
\end{eqnarray}

Following \cite{Chatzis:2025dnu} we note that in the limit $\ell\to 0$
the Coulomb branch metric in eq.(\ref{S5-1}) reduces to  that in eq.(\ref{ARmetric}). When we set a finite (non-zero) value for $\ell$, we are studying an interesting deformation of the result in the previous section.
For the present case (Coulomb branch with gapped IR) the expressions of the separations in eq.(\ref{separationsTY}) and the entanglement in eq.(\ref{regulatedEE}) read,
\begin{eqnarray}
& & \frac{T}{2 c_t L^4}=\frac{Y}{2 c_y L^4}= \int_{u_0}^\infty du~ \frac{u}{\sqrt{\left( u^6 +\epsilon \ell^2 u^4 - L^6q^6\right) \left[(u^6-u_0^6) + \epsilon \ell^2 (u^4-u_0^4) \right]}} .\label{YT-separationcoulomb}   
\end{eqnarray}
For the regularised entanglement entropy, we have
\begin{eqnarray}
& &\frac{2 G_{10}}{{\cal N}_{II}}~S_{EE}=\int_{u_0}^\infty du ~ u \sqrt{\frac{\left( u^6 +\epsilon \ell^2 u^4 - L^6q^6\right)}{\left[(u^6-u_0^6) + \epsilon \ell^2 (u^4-u_0^4) \right]}} - \int_{u_*}^\infty u ~ du.\label{SEEcoulomb}    
\end{eqnarray}
We perform the change of variables,
\begin{equation*}
u=\frac{u_0}{r}, ~~\text{and define}~~\mu=\frac{\epsilon\ell^2}{u_0^2},~~\nu=\frac{L^6q^6}{u_0^6}.    
\end{equation*}
After this,  we have
\begin{eqnarray}
& &  \frac{T~ u_0^4}{2 c_t~L^4}=\frac{Y~u_0^4}{2 c_y~L^4}= \int_0^1 dr ~\frac{r^3}{\sqrt{\left( 1+\mu~ r^2 -\nu ~r^6\right) \left[(1-r^6) +\mu ~r^2 (1-r^4) \right]}}. \label{separationscoulombbranch}\\
& & \frac{2 G_{10}  }{{\cal N}_{II} u_0^2} ~ S_{EE}= \int_0^1 \frac{dr}{r^3}\sqrt{\frac{\left( 1+\mu~ r^2 -\nu ~r^6\right)}{\left[(1-r^6) +\mu ~r^2 (1-r^4) \right]}} - \int_0^{\frac{u_0}{u_*}} \frac{dr}{r^3}.\label{SEEcoulombbranch}
\end{eqnarray}

\begin{figure}
    \centering
    \includegraphics[width=0.5\linewidth]{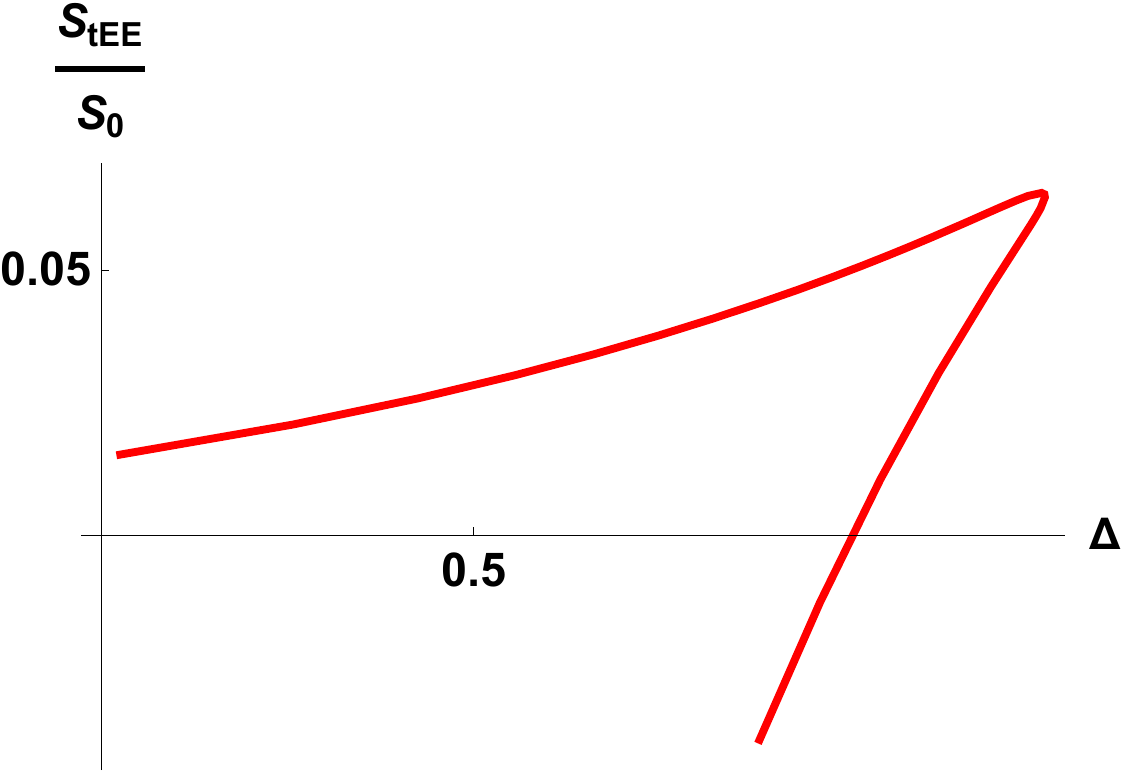}
    \caption{Plot for tEE vs system size where we set $\epsilon=1$, $q=1$, $L=1$ and $\ell=0.001$. An almost identical plot can be obtained for the choice $\epsilon =-1$.}
    \label{fig:placeholder}
\end{figure}

Let us combine \eqref{separationscoulombbranch} and \eqref{SEEcoulombbranch} to find the separation
\begin{align}
    \Delta^2=Y^2-T^2 =\frac{4L^2}{u_0^2}(1+\mu -\nu)\mathcal{J}_1^2.\label{armando}
\end{align}
where $\mathcal{J}_1$ is the integral in \eqref{separationscoulombbranch}. We used eq.(\ref{turningdiego}) below to derive the above expression.

The integrals in  eqs.(\ref{separationscoulombbranch})-(\ref{SEEcoulombbranch}) need to be numerically evaluated. Notice that in the limit $\ell\to 0$ (or $\mu\to 0$), these expressions above reproduce those in the Anabal\'on-Ross model, see eqs.(\ref{definitionsconf}),(\ref{TY-confining}). As before the EE is regulated (the divergent parts coming from the lower limit of integration in eq.(\ref{SEEcoulombbranch}) do cancel). Also, note that the separations $T,Y$ and the entanglement depend on the turning point $u_0$ both explicitly, and implicitly, as the parameters $(\mu,\nu)$ are themselves functions of $u_0$.
\\
The turning point is obtained by solving
\begin{equation}
F(u_0)^2=c_y^2-c_t^2,~~\longrightarrow~~u_0^6+\epsilon \ell^2 u_0^4= L^6(q^6+ c_y^2-c_t^2). \label{turningdiego}   
\end{equation}
The solution is,
\begin{eqnarray}
& & 27 u_0^6=\Bigg[-\epsilon \ell^2 +\frac{\ell^4}{\cal Z}+{\cal Z}\Bigg]^3,\label{equ0ANO}\\
& & \text{where}~~ {\cal Z}^3=\frac{27 m}{2}\left(1-\frac{2\epsilon \ell^6}{27 m}+\sqrt{1-\frac{4\epsilon \ell^6}{27 m}} \right), ~\text{with}~ m= L^6\left(q^6+c_y^2-c_t^2 \right).\nonumber
\end{eqnarray}
As observed above, note that for $\ell\to 0$, the turning point is that of eq.(\ref{u0conf}).
\\
In the case of the Anabal\'on-Ross model, we observe below eq.(\ref{u0conf}), that the confinement scale (represented there by the parameter $Q$), allows for the existence of Type I surfaces (with real-valued turning point $u_0$) even in the purely time-like case $c_y=0$. In the case of the model by Anabal\'on, Nastase and Oyarzo we have two parameters $q$ and $\ell$, and we also have the two choices of sign $\epsilon=\pm 1$. In the case $\epsilon=-1$, exploring the Coulomb branch with the parameter $\ell$ makes the effect above (Type I/real embeddings even for purely time-like slabs in the field theory)  more prevalent, provided $m>0$. The opposite occurs for the branch of solutions $\epsilon=+1$.


\subsubsection{Gapped linear quivers and universality of the entanglement entropy}
In this section we present two metrics and dilaton fields (as above, we do not quote the Ramond and other Neveu-Schwarz fields). These backgrounds were written and studied in the papers \cite{Chatzis:2024kdu,Chatzis:2024top,Chatzis:2025dnu}, we refer the reader to these papers for details.
For our purposes, it is useful to think about the backgrounds here presented as 'embellishments' of the Anabal\'on-Ross and Anabal\'on-Nastase-Oyarzo models in Sections \ref{sectionAR} and \ref{coulombbranchgapped}. Of course, the added 'ornament' essentially changes the physical interpretation. In a nutshell, the dynamics of the models in Sections \ref{sectionAR}, \ref{coulombbranchgapped} is now translated to linear quiver field theories, like the ones we studied in Section \ref{sectionexamplescft}.

The four dimensional linear quiver version of the Anabal\'on-Ross model is given by a background in ten-dimensional IIA-supergravity whose metric and dilaton read,
\begin{equation}
\label{eq:N=2} 
\begin{aligned}
\mathrm{d}s^2= f_1^{\frac{3}{2}} f_5^{\frac{1}{2}}\bigg[4\mathrm{d}s^2_5+f_2 D\mu_iD\mu_i+f_4(\mathrm{d}\sigma^2+\mathrm{d}\eta^2)+f_3 (\mathrm{d}{\chi} +\mathcal{A})^2\bigg],~~~
e^{\frac{4}{3}\Phi}=  f_1 f_5
\end{aligned}
\end{equation}
We have defined,
\begin{eqnarray} 
& &\mathrm{d}s^2_{5}=\frac{u^2}{l^2} (\lambda\mathrm{d}t^2+\mathrm{d}y^2 + \mathrm{d}x^2 +  f(u)\mathrm{d}\phi^2) + \frac{l^2 ~\mathrm{d}u^2}{ u^2 f(u)} .\label{ds^2AR}\nonumber\\
& &  f(u) = 1  - \frac{Q^2l^2}{u^6},~~  \mathcal{A} =Q \left( \frac{1}{u^2}- \frac{1}{u _{*}^2}\right)\mathrm{d}\phi \,,~~u_*= (Q l)^{1/3}.\label{ARgaugefield}\nonumber\\
& &\mu_1= \sin\theta \sin\varphi,~\mu_2= \sin\theta \cos\varphi, ~\mu_3= \cos\theta.\label{mui}\nonumber\\
& & D \mu_1
=\mathrm{d} \mu_1+2 \mu_2 \mathcal{A}, ~~D \mu_2
=\mathrm{d} \mu_2-2\mu_1 \mathcal{A}, ~~ D \mu_3=\mathrm{d} \mu_3,\nonumber\\
& &  f_1=\bigg(\frac{\dot{V}\tilde{\Delta}}{2V''}\bigg)^{\frac{1}{3}},~~~~f_2 = \frac{2V''\dot{V}}{\tilde{\Delta}},~~~~f_3=\frac{4\sigma^2}{\Lambda},~~~~f_4 = \frac{2V''}{\dot{V}},\nonumber\\
& & f_5=\frac{2\Lambda V''}{\dot{V}\tilde{\Delta}},
    ~~~~~\tilde{\Delta} =\Lambda(V'')^2+(\dot{V}')^2,~~~~~\Lambda=\frac{2\dot{V}-\ddot{V}}{V''}.\label{linearquiverAR}
 \end{eqnarray} 
With these definitions, we can calculate the entanglement entropy. We choose as in Section \ref{sectionAR}, the eight manifold
 $\Sigma_8 = [x, \phi, u,\sigma,\eta,\theta,\varphi,\chi]$, with $t = t(u)$ and $y = y(u)$. The induced metric on $\Sigma_8$ is
\begin{eqnarray}
ds^2_8 = & & \left(f_1^3 f_5\right)^{\frac{1}{2}}\Bigg(4\left[\frac{u^2}{l^2}(\lambda t'^2 + y'^2) + \frac{l^2}{u^2 f(u)}\right] du^2 + \frac{4u^2}{l^2} dx^2 + \frac{4u^2}{l^2} f(u) d\phi^2 + \nonumber\\
& &~~~~~~~~~~~~~~f_2 D\mu_iD\mu_i+f_4(\mathrm{d}\sigma^2+\mathrm{d}\eta^2)+f_3 (\mathrm{d}{\chi} +\mathcal{A})^2\Bigg).
\end{eqnarray}
As an intermediate step we calculate,
\begin{equation}
e^{-4\Phi}\det[g_{\Sigma_8}]= f_1^9 f_2^2 f_4^2 f_3 f_5 \text{Vol}_{S^2(\theta,\varphi)}\text{Vol}_{S^1(\chi)} \left[\frac{u^2}{l^2}+ \frac{u^6}{l^6}f(u)(y'^2+\lambda t'^2 )\right].    
\end{equation}
Using this,
the entanglement entropy is given by
\begin{eqnarray}
& & S^{(\lambda)}_{EE} = \frac{\tilde{\mathcal{N}}}{4G_{10}} \int_{u_0}^\infty \sqrt{G^2(u) + F^2(u)(\lambda t'^2(u) + y'^2(u))}, \label{EEconflinear1} \\
& & G(u) = \frac{u}{l}, \quad F(u) = \frac{u^3}{l^3} \sqrt{f(u)}, \quad \tilde{\mathcal{N}} =256 \pi^2 L_x L_\phi  \int_{0}^{\infty} d\sigma \int_0^Pd\eta ~\sigma ~\dot{V}~V'', \quad L_\phi = \frac{1}{3Q}. \nonumber
\end{eqnarray}
There are a couple of interesting observations to make:
\begin{itemize}
    \item{First, notice that the functions $G(u), F(u)$ characterising the surface that calculates the EE are the same as those in the Anabal\'on-Ross model. In fact, compare eqs.(\ref{EEconf}) and (\ref{EEconflinear1}). This implies that the dynamics of the eight surface, the separations in $T$ and $Y$ and the expression of the EE in terms of them are exactly those in eqs.(\ref{TY-confining}),(\ref{SDeltaconf}). Note that the prefactor $\hat{\cal N}$ in eq.(\ref{EEconf}) differs from  $\tilde{\cal N}$ in eq.(\ref{EEconflinear1}).}
    \item{The second observation is that the coefficient ${\tilde{\cal N}}$ is the same as that encountered when discussing four dimensional linear quiver SCFTs, see eq.(\ref{calNparaGM}). }
    \end{itemize}
In the parlance of \cite{Chatzis:2025dnu}, the entanglement entropy is a 'universal' observable. These are observables for which the information coming from the flow from conformal to gapped QFT (the functions $F(u),G(u)$ in this case), separates (or factorises) from the information of the UV-CFT (the coefficient $\tilde{\cal N}$ in this case). As explained in \cite{Chatzis:2025dnu} this is a consequence of the conjecture posed by Gauntlett and Varela \cite{Gauntlett:2007ma}, proven in \cite{Cassani:2019vcl}.

Briefly, we quote the result of translating to  four dimensional linear quivers the Coulomb branch plus gap in the QFT dual to the background in eq.(\ref{S5-1}). The reader should appreciate that even when the system becomes quite involved (the functions $\tilde{f}_i(u,\sigma,\eta)$ are not factorised), after the calculation is done, we arrive at the same result in eqs.(\ref{EEforANO})-(\ref{FGCoulomb}), with the difference appearing in the coefficient ${\cal N}_{II}$ related to the UV-CFT.

In fact, the background metric and dilaton are written in \cite{Chatzis:2025dnu} for an infinite family of linear quiver UV-CFTs that explore the Coulomb branch ending with a mass gap. The background metric and dilaton are,

\begin{eqnarray}
&         &\mathrm{d}s^2_{10}= \tilde{f}_1^{\frac{3}{2}} \tilde{f}_5^{\frac{1}{2}} \left[ 4\tilde{\gamma} \mathrm{d}s^2_5 + \tilde{f}_2 \mathrm{D}\mu_i\mathrm{D}\mu^i + \tilde{f}_3(\mathrm{d}\chi+B)^2 + \tilde{f}_4(\mathrm{d}\sigma^2 + \mathrm{d}\eta^2) \right],\\
&        &e^{\frac{4}{3}\Phi} = \tilde{f}_1\tilde{f}_5. \nonumber\end{eqnarray}
The functions $\tilde{f}_i(u,\sigma,\eta)$ read
\begin{equation}\label{IIA_functions}
    \begin{split}
        & \tilde{f}_1=\left(\frac{\dot{V}\tilde{\Delta}}{2V^{\prime \prime}}\right)^{1/3},\quad \tilde{\gamma} = \frac{ Z}{X(u)}=\frac{Z}{\lambda^2(u)},\quad \tilde{f}_2=\frac{2\dot{V}V^{\prime\prime}}{Z^2\tilde{\Delta}}, \quad \tilde{f}_3 = \frac{4X^3\sigma^2V^{\prime\prime}}{2X^3\dot{V}-\ddot{V}}Z \, ,
        \\
        &\tilde{f}_4=\frac{2V^{\prime\prime}}{\dot{V}}Z,\quad \tilde{f}_5 = \frac{2(2X^3\dot{V}-\ddot{V})}{Z^2\dot{V}\tilde{\Delta}},\\
        &
        \tilde{\Delta}=(\dot{V}^{\prime})^2+V^{\prime\prime}(2\dot{V}-\ddot{V}) ,\quad Z=\left[ \frac{(\dot{V}^{\prime})^2+V^{\prime\prime}(2X^3\dot{V}-\ddot{V})}{(\dot{V}^{\prime})^2+V^{\prime\prime}(2\dot{V}-\ddot{V})}\right]^{1/3}.
    \end{split}
\end{equation}
We also defined,
\begin{eqnarray}
& & ds_5^2=\frac{u^2\lambda(u)^2}{L^2} \left(-dt^2+ dy^2+ dx^2+ L^2 \tilde{f}(u) d\phi^2 \right) +\frac{du^2}{u^2\lambda(u)^{{4}} \tilde{f}(u)},
\nonumber\\[5pt]
& & \mu_1= \sin\theta \cos\varphi,~~~~~~~\mu_2=\sin\theta \sin\varphi,~~~~~~~\mu_3= \cos\theta,
\nonumber\\[5pt]
& &D\mu_i= \left(d\mu_1+ 2\mu_2 A_{1\phi} d\phi\right)\delta_{i,1}+\left(d\mu_2 -2\mu_1 A_{1\phi}d\phi \right)\delta_{i,2}+ d\mu_3 \delta_{i,3}.\label{betoalonso}
\end{eqnarray}
The one-forms $A_{1}, A_2, A_3=B$ are those in eq.(\ref{one-forms-ANO}) and $\tilde{f}(u),\lambda(u)$ are defined in eq.(\ref{functionsANO}).
Choosing the eight manifold $\Sigma_8=[w,\phi,u,\theta,\varphi,\chi,\sigma,\eta]$, one finds,
\begin{equation}
e^{-4\Phi}\det[g_{\Sigma_8}]= 4^3 \tilde{f}_1^9\tilde{f}_2^2\tilde{f}_4^2\tilde{f}_3 \tilde{f}_5 \tilde{\gamma}^3\left[\frac{u^2}{L^2}+ \frac{u^6 \lambda^6(u) \tilde{f}(u)}{L^4}(y'^2-t'^2) \right]  \text{Vol}_{S^2(\theta,\varphi)}\text{Vol}_{S^1(\chi)}  .
\end{equation}
Here, we see the conjecture of Gauntlett and Varela \cite{Gauntlett:2007ma} at work. In fact, there is a factorisation of the $(\sigma,\eta)$-dependent part from the $u$-dependent one. This is non-trivial, since the functions $\tilde{f}_i(u,\sigma,\eta)$ depend on their variables in a non-factorised fashion. A lengthy but straightforward calculation gives expressions like those in eqs.(\ref{EEforANO})-(\ref{FGCoulomb}). The constant ${\cal N}_{II}$ is now replaced by
\begin{equation*}
 \tilde{\cal N}_{II}= 256\pi^2 L_x L_\phi  \int_{0}^{\infty} d\sigma \int_0^Pd\eta ~\sigma ~\dot{V}~V'' .  
\end{equation*}
The observations about universality itemised below eq.(\ref{EEconflinear1}) are applicable to this case. 
\\
Let us now study backgrounds in class III.

\subsection{Backgrounds of class III}
In this section we discuss the entanglement entropy in the case of some backgrounds in the class III classification. We consider background metric and dilaton $\Phi$ of the form,
\begin{eqnarray}
& & ds^2_{st}=e^{\Phi} \hat{h}^{-\frac{1}{2}}(u)\Bigg[\lambda dt^2 +  dy^2 +dx_1^2+ dx_2^2 + \hat{h}(u) e^{2k(u)} du^2+ \hat{h}(u) e^{2h(u)}(d\theta^2+\sin^2\theta d\varphi^2) \nonumber\\
& & ~~~~~~~~~~~~~~~~~~~~~~~~+\frac{\hat{h}(u) e^{2g(u)}}{4}\left[\left(\omega_1-A^{(1)}\right)^2+\left(\omega_2-A^{(2)}\right)^2\right]+
 \frac{\hat{h}(u) e^{2k(u)}}{4}\left(\omega_3-A^{(3)}\right)^2\Bigg],\nonumber\\
 & &\Phi=\Phi(u).\label{BBmetric}
\end{eqnarray}
In the equation above, we have defined
\begin{eqnarray}
& &\omega_1= \cos\tilde{\psi} d\tilde{\theta}+\sin\tilde{\psi} \sin\tilde{\theta}d\tilde{\varphi}, ~\omega_2= -\sin\tilde{\psi} d\tilde{\theta}+\cos\tilde{\psi} \sin\tilde{\theta}d\tilde{\varphi},~\omega_3=d\tilde{\psi}+\cos\tilde{\theta}d\tilde{\varphi},\nonumber\\
& &A^{(1)}= -a(u) d\theta, ~~~~A^{(2)}=a(u) \sin\theta d\varphi,~~~~A^{(3)}=-\sin\theta d\varphi,\label{one forms}
\end{eqnarray}
and the functions,
\begin{eqnarray}
& & e^{2h}=\frac{(P^2-Q^2)}{4\left(P \coth(2u) -Q\right)},~~~~e^{2g}= \left(P \coth(2u) -Q\right), ~~~~ e^{2k}= \frac{P'}{2}, ~~ e^{4\Phi}= 2e^{4\Phi_0}\frac{\sinh^2(2u)}{(P^2-Q^2)P'},\nonumber\\
& &a= \frac{P}{P\cosh(2u) -Q\sinh(2u)},~~~\hat{h}= 1-\kappa^2 e^{2\Phi},~~~Q= N_c\Big(2 u \coth (2u)-1 \Big).\label{functionsBB}
\end{eqnarray}
The function $P(u)$ satisfies the ordinary differential equation,
\begin{equation}
P''+P'\Bigg(\frac{P'-Q'}{P+Q}+\frac{P'+Q'}{P-Q}-4\coth(2u) \Bigg)=0.\label{mastereq}    
\end{equation}
This metric and dilaton should be complemented by Ramond and Neveu-Schwarz fields, we direct the reader to \cite{Conde:2011aa} for a careful explanation of the system and different solutions of the ODE (\ref{mastereq}). The backgrounds of the form given by eqs.(\ref{BBmetric})-(\ref{mastereq}) encode various known supergravity duals:
\begin{itemize}
\item{When the parameter $\kappa=0$ (appearing in the definition of $\hat{h}(u)$), we describe systems of $N_c$ D5 branes wrapping a two cycle inside the resolved conifold \cite{Maldacena:2000yy}.}
\item{For nonzero $\kappa$ (and the function $P(u)$ found in \cite{Gaillard:2010qg, Elander:2011mh,Hoyos-Badajoz:2008znk}), we describe the Baryonic Branch \cite{Butti:2004pk} of the Klebanov-Strassler solution \cite{Klebanov:2000hb}. For generic values of the constant $\kappa$, the field theory is coupled to gravity, whilst for $\kappa=e^{-\Phi(\infty)}$, we describe the solution of \cite{Butti:2004pk}, decoupled from gravity.}
\end{itemize}
Various papers have elaborated on solutions of this type, connecting and generalising them in different ways \cite{Maldacena:2009mw, Caceres:2011zn, Gaillard:2010qg, Elander:2011mh, Nunez:2008wi, Elander:2009pk}.
We are interested in two kinds of solutions to the ODE in eq.(\ref{mastereq}). One kind of solution  is exact,
\begin{equation}
 P= 2 N_c~ u.\label{exactmaster}
\end{equation}
This exact solution is only acceptable if the constant $\kappa=0$, making $\hat{h}(u)=1$. When plugged into the configuration, we find the exact non-singular solution corresponding to D5 branes wrapping a two cycle inside the resolved conifold \cite{Chamseddine:1997mc, Chamseddine:1997nm, Maldacena:2000yy}.

\begin{figure}[h]
\centering
\includegraphics[width=0.8\linewidth]{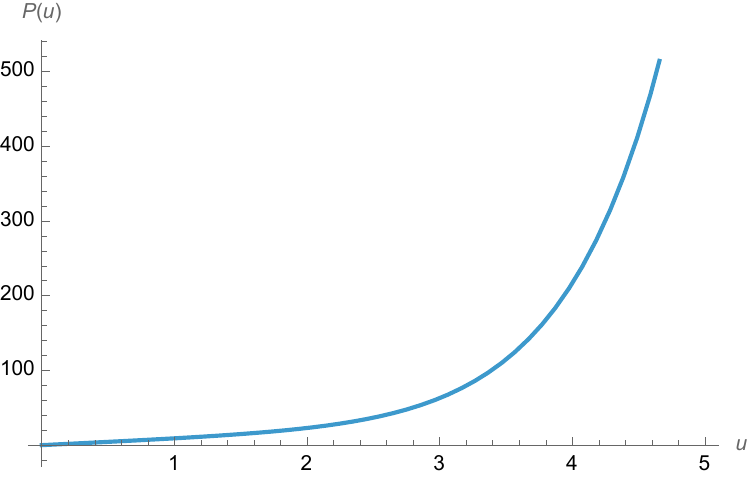}
\caption{The function $P(u)$ interpolating between the UV and IR expansions in eqs.(\ref{UV-II-N.A}),(\ref{P-IR.A}).}
\label{Pofu}
\end{figure}

\begin{figure}[h]
\centering
\includegraphics[width=0.8\linewidth]{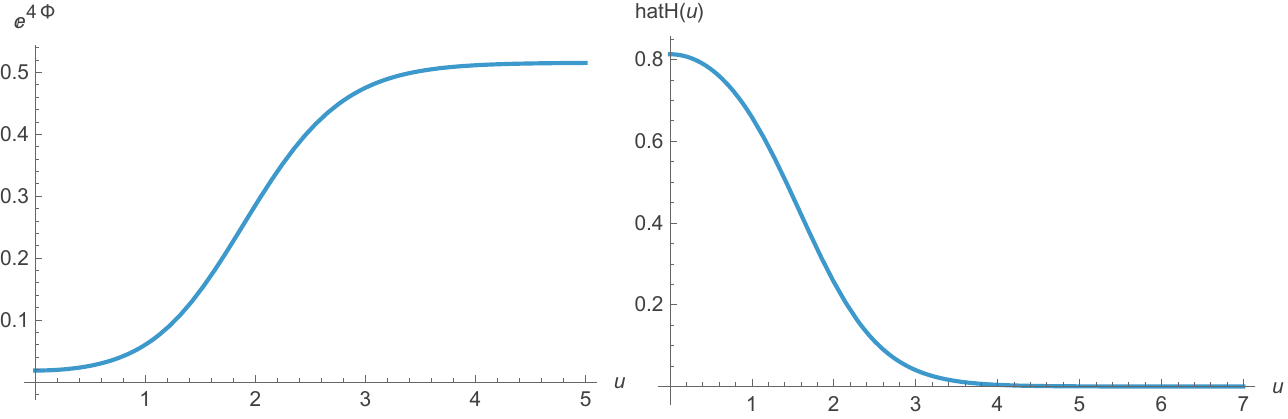}
\caption{The functions $e^{4\Phi(u)}$ and $\hat{h}(u)$ computed using $P(u)$ in Figure \ref{Pofu}. We chose $\Phi_0=0$ and $\kappa=e^{-2\Phi(\infty)}$.}
\label{Phihathtuned}
\end{figure}

The other kind of solution is only known numerically. In a series expansion for large values of the $u$-coordinate (in the UV) we find the expression
\begin{eqnarray}
& &P=e^{\frac{4u}{3}}\Big[ c_+  
+\frac{e^{-\frac{8u}{3}} N_c^2}{c_+}\left(
4u^2 - 4u +\frac{13}{4} \right)+ e^{-4u}\left(
c_- -\frac{8c_+}{3}u \right)+\nonumber\\
& & 
~~~~~~~~~~~~~~~ + \frac{N_c^4 e^{-\frac{16u}{3}}}{c_+^3}
\left(\frac{18567}{512}+\frac{2781}{32}u +\frac{27}{4}u^2 +36u^3\right)  + O(e^{-\frac{20u}{3}})
\Big]\,.
\label{UV-II-N.A}
\end{eqnarray}
The reader can check that the geometry  in eq.(\ref{BBmetric}) asymptotes to the conifold 
after using the expansion in eq.(\ref{UV-II-N.A}).
\\
In the IR (that is, close to the origin of the space that we take to be $u=0$), we find a solution of eq.(\ref{mastereq}) in an expansion for small-$u$,
\begin{equation}
P= h_1 u+ \frac{4 h_1}{15}\left(1-\frac{4 N_c^2}{h_1^2}\right)u^3
+\frac{16 h_1}{525}\left(1-\frac{4N_c^2}{3h_1^2}-
\frac{32N_c^4}{3h_1^4}\right)u^5+O(u^7)\,.
\label{P-IR.A}
\end{equation}
For $h_1= 2N_c+\zeta$, being $\zeta>0$ some positive number. Note that for $h_1=2N_c$ we are back in the exact solution of eq.(\ref{exactmaster}).
This space is free of singularities as can be checked by computing curvature invariants. Figure \ref{Pofu} shows the function $P(u)$ for the numerical solution interpolating between the expansions in eqs.(\ref{UV-II-N.A}) for large $u$ and (\ref{P-IR.A}) for small $u$. On the other hand, Figure \ref{Phihathtuned} shows $e^{4\Phi(u)}$ and $\hat{h}(u)$ for $\Phi_0=0$ and $\kappa=e^{-2\Phi(\infty)}$.
It should be emphasised that for both the exact solution (with parameter $\kappa=0$) and the numerical one (with parameter $\kappa=e^{-\Phi(\infty)}$), the calculation of Wilson loops results in an area law, namely the dual QFTs are {\it confining}. We could expect that the time-like entanglement entropy leads to a phase transition when plotted against the separation of the slab (as it happens with the confining models of Section \ref{sectionAR} and \ref{coulombbranchgapped}). This expectation is not borne-out. As we discuss the link between phase transitions in the tEE and the confining character of the QFT must be refined by the requirement that the UV system is a local field theory. Let us discuss this in detail.
\\

\begin{figure}[h]
\centering
\includegraphics[width=0.8\linewidth]{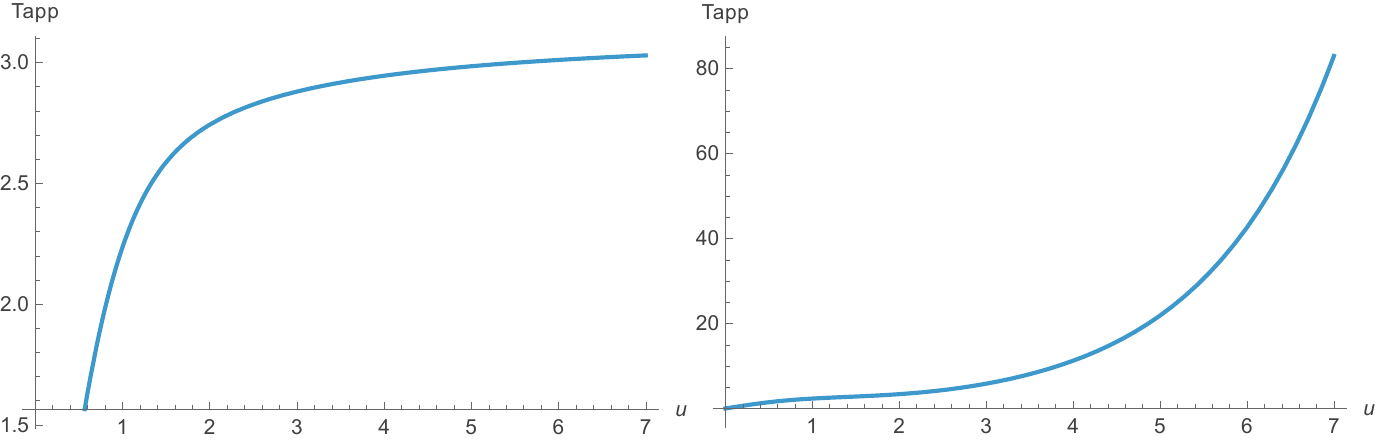}
\caption{The function $T_{app}$. On the left, for the exact solution $P=2N_c u$. On the right, for a de-tuned case ($\kappa<e^{\Phi(\infty)}$). In both cases there is no double valuedness, hence, no phase transition. }
\label{Tappexactotra}
\end{figure}

Let us now compute the entanglement entropy. We choose the eight manifold $\Sigma_8=[x_1, x_2, u, \theta,\varphi, \tilde{\theta},\tilde{\varphi},\tilde{\psi} ]$ with $t(u),y(u)$. We calculate,
\begin{eqnarray}
& &ds^2_{\Sigma_8}=     
e^{\Phi} \hat{h}^{-\frac{1}{2}}(u)\Bigg[  dx_1^2+ dx_2^2 + \left(\hat{h}(u) e^{2k(u)} +\lambda t'^2+y'^2 \right) du^2+ \hat{h}(u) e^{2h(u)}(d\theta^2+\sin^2\theta d\varphi^2) \nonumber\\
& & ~~~~~~~~~~~~~~~~~~~~~~~~+\frac{\hat{h}(u) e^{2g(u)}}{4}\left[\left(\omega_1-A^{(1)}\right)^2+\left(\omega_2-A^{(2)}\right)^2\right]+
 \frac{\hat{h}(u) e^{2k(u)}}{4}\left(\omega_3-A^{(3)}\right)^2\Bigg],\nonumber\\
 & &e^{-4\Phi}\det[g_{\Sigma_8}]=\frac{\hat{h}^2}{64}e^{4\Phi+4h+4g+4k}\left(\frac{e^{-2k}}{\hat{h}} (\lambda t'^2+y'^2)+1 \right) \text{Vol}_{S^2}\text{Vol}_{S^3}\nonumber\\
 & &4 G_{10}S_{EE}= {\cal N}_{III} \int du \sqrt{G^2(u)+ F^2(u) (y'^2+\lambda t'^2)},\label{SEEBB}\\
 & & G^2(u)= e^{4\Phi+4h+4g+4k}\hat{h}^2= \frac{e^{4\Phi_0}}{32}\sinh^2(2u) (P^2-Q^2)P'\hat{h}^2,\nonumber\\
 & &F^2(u)=\frac{e^{-2k} G^2(u)}{\hat{h}}=\frac{e^{4\Phi_0}}{16}\sinh^2(2u) (P^2-Q^2)\hat{h} .\nonumber
\end{eqnarray}
In summary, we have two different solutions and want to compute the time-like EE for each of them. For both solutions the integrals needed cannot be evaluated exactly and a demanding numerical analysis is needed (which is not the objective of this work). Instead, we study the approximate expressions, in particular that for the approximate time separation $T_{app}$ in eq.(\ref{Tapp}). We plot $T_{app}$ obtained from the exact solution $P=2 N_c u$ in eq.(\ref{exactmaster}). We find that as a function of $u_0$, it  is monotonically increasing, see left insert of Figure \ref{Tappexactotra}. The maximum value of $T_{app}$ in the background of N$_c$ D5 branes is related to the scale inherent to the Little String Theory. In this case there is no phase transition in the tEE.

\begin{figure}[h]
\centering
\includegraphics[width=0.8\linewidth]{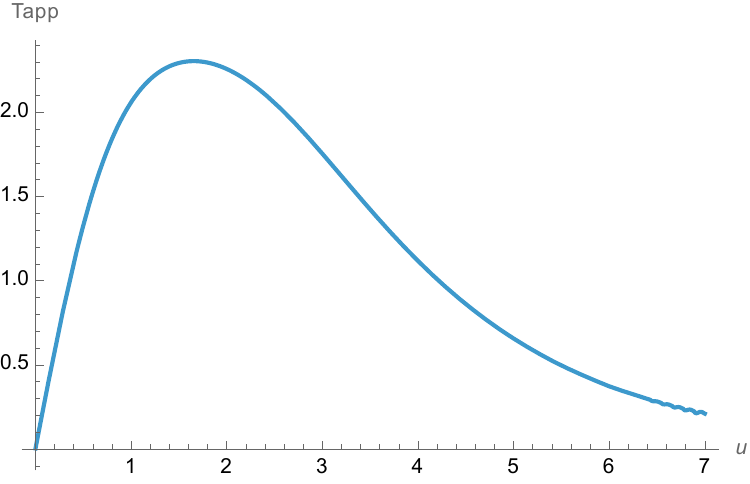}
\caption{The approximate separation $T_{app}$ calculated with $P(u)$ interpolating between the UV and IR expansions in eqs.(\ref{UV-II-N.A}),(\ref{P-IR.A}) and for the finely tuned case $\kappa=e^{\Phi(\infty)}$.}
\label{TappBB}
\end{figure}

We move to study the same $T_{app}$ for the numerical solution in eqs.(\ref{UV-II-N.A})-(\ref{P-IR.A}). Remember that in this case we have a free parameter $\kappa$. If we choose $\kappa^2< e^{-2\Phi(\infty)}$, we find an analogous situation. In fact, the $T_{app}$ is monotonically increasing (and diverges for $u_0\to\infty$), and no phase transition for the tEE is encountered, see the right insert in Figure \ref{Tappexactotra}.
\\
Finally for the particularly tuned case $\kappa^2=e^{-2\Phi(\infty)}$, we find a double-valued $T_{app}$, signaling the presence of a phase transition in the tEE, see Figure \ref{TappBB}.

Let us explain physically these behaviours. In the paper \cite{Klebanov:2007ws} it was proposed that an alternative indicator of confinement is the existence of a phase transition in the (space-like) entanglement entropy. This was critically analysed in \cite{Kol:2014nqa} and later in \cite{Jokela:2020wgs}. The result is that the phase transition in the EE indicates confinement only if the UV of the holographic QFT is {\it local}.
Similar proposals were made for the time-like entanglement entropy (as a tool to diagnose confinement), see for example \cite{Afrasiar:2024lsi, Chu:2019uoh, Nunez:2025gxq}.
We find that a similar caveat should be in place for the tEE case:
\\{\it the phase transition of the tEE is an indicator of confinement if and only if the UV of the holographic QFT is local}.

The exact solution of eq.(\ref{exactmaster}) leads to a UV system represented by $N_c$ D5 branes, hence a Little String theory (non-local). This implies that the phase transition in the tEE (double valuedness in $T_{app}$ or $T$) is not expected. By the same token, the system represented by the approximate solution, is only field theoretical when $\kappa^2=e^{-2\Phi(\infty)}$. Otherwise, the dual field theory is coupled to gravity (10 dim strings) and non-local. An explanation of the importance of the finely-tuned value $\kappa^2=e^{-2\Phi(\infty)}$ is given in the papers \cite{Maldacena:2009mw, Caceres:2011zn, Elander:2011mh}. Briefly, this special value 'switches-off' an irrelevant operator on the QFT (the baryonic branch of the Klebanov-Strassler field theory). We expect similar behaviour in the newly presented background of \cite{Macpherson:2025pqi}.

We close this section here and move to final comments and conclusions.
\section{Conclusions  and future directions}\label{sectioncoclusions}
Let us start with a summary, to then provide some possible future lines of research.

This paper explores a novel approach to computing time-like entanglement entropy (tEE) in holographic settings, emphasizing the similarities with the formalism developed when studying  Wilson lines and the functionals that have been proposed
to capture its behavior. 

We propose a refined formalism for spacetime entanglement entropy, particularly tailored to slab regions, and introduce practical approximations for time-separation and the entanglement itself. 

An interesting contribution is the derivation of expressions for the entanglement entropy across three classes of ten-dimensional supergravity backgrounds, with clear applicability to both confining gauge theories and general conformal field theories (CFTs). Importantly, we offer a method to compute the Liu-Mezei central charge using real-time techniques. The treatment is grounded in ten-dimensional top-down models and accommodates extensions to eleven-dimensional supergravity. The paper addresses some of the subtleties of complex extremal surfaces and their physical significance, providing theoretical insights and practical tools to advance the understanding of holographic entanglement. This work  delivers a conceptual and computational framework for tackling real-time entanglement in curved spacetimes.
\\
Some possible new lines of investigation that this paper opens are:
\begin{itemize}
\item{The analysis focuses mainly on slab regions, a logical next step is to generalise the time-like entanglement entropy (tEE) framework to spherical or arbitrary entangling surfaces (in the case of confining models). This would broaden the applicability of the methods and deepen understanding of spacetime entanglement geometry.}
\item{While the role of the Ramond and Neveu-Schwarz fields is absent in our treatment, it would be interesting to know if this sector of the string background contains (or shares) some of the information that the EE contains.}
\item{
A more careful 
implementation of the  numerical simulations of our proposed expressions to compare approximated tEE with exact holographic computations, especially for complex extremal surfaces. This could validate or refine our approximations.}
\item{ It would be nice to extend our analysis
to thermal states or  other charged backgrounds. This may reveal how time-like entanglement encodes thermodynamic or hydrodynamic information in the dual theory.}
\item{It might be interesting to extend our calculations to holographic duals to 1d CFTs and Matrix models, for example those in \cite{Lin:2005nh, Donos:2010va, Lozano:2017ole, Komatsu:2024bop}. }
\item{
Finally, the addition of quantum corrections (either in $g_s$ or $1/N$) in the bulk or subleading contributions to tEE could improve the precision of holographic predictions.}
\end{itemize}
We hope to report on these and other issues in the future.
\section*{Acknowledgments:} We are happy to thank various colleagues for discussions in the recet months, that enriched our recent works on this topic. For this, we are grateful to: Dimitrios Giataganas, Wu-zhong Guo, Michal Heller, Fabio Ori, Simon Ross, Alexandre Serantes, Tadashi Takayanagi and Jonathan Whittle.
\\
DR would like to acknowledge The Royal Society, UK for financial assistance. DR also acknowl- edges the Mathematical Research Impact Centric Support (MATRICS) grant (MTR/2023/000005) received from ANRF, India. C. N. is supported by STFC’s grants ST/Y509644-1, ST/X000648/1 and ST/T000813/1.

\bibliographystyle{JHEP}
\bibliography{main.bib}

\end{document}